\documentclass[aps,prc,twocolumn,showpacs,floatfix,preprintnumbers,%
superscriptaddress,amsmath,amssymb,longbibliography]{revtex4-1}

\usepackage{tabularx}
\usepackage{mathtools}
\usepackage{pifont}
\usepackage[colorlinks,citecolor=blue]{hyperref}
\usepackage[colorinlistoftodos]{todonotes}
\usepackage[normalem]{ulem}
\usepackage{color}
\usepackage[utf8]{inputenc}
\usepackage{bm}
\usepackage{caption}
\usepackage{subcaption}
\usepackage{booktabs}

\makeatletter
\def\@bibdataout@aps{%
\immediate\write\@bibdataout{%
@CONTROL{%
apsrev41Control%
\longbibliography@sw{%
    ,author="08",editor="1",pages="1",title="0",year="1"%
    }{%
    ,author="08",editor="1",pages="1",title="",year="1"%
    }%
  }%
}%
\if@filesw \immediate \write \@auxout {\string \citation {apsrev41Control}}\fi 
}
\makeatother

\graphicspath{{./}{figures/}}

\definecolor{pastelgray}{rgb}{0.81, 0.81, 0.77}
\definecolor{beaublue}{rgb}{0.9, 0.9, 0.93}

\begin{document}
\title{Equation-of-State Independent Relations in Rapidly Rotating Hybrid Stars}

\author{Sujan Kumar Roy}
\affiliation{Physics Group, Variable Energy Cyclotron Centre, Kolkata 700064, India }
\date{\today}

\begin{abstract}
We investigate 22 hadronic equations of state that incorporate the possibility of heavy baryon formation at sufficiently high densities, with the aim of establishing quasi-universal relations for both slowly and rapidly rotating neutron stars. The selected equations of state satisfy current observational constraints, such as those from NICER and GW170817. Our fitting results yield relations between various macroscopic quantities that are approximately independent of the underlying equation of state, with typical deviations on the order of $\mathcal{O}$(10\%) for neutron stars containing heavy baryonic degrees of freedom. The approximately universal I-Love-Q relations for slowly rotating neutron stars and the I-C-Q relations for rapidly rotating configurations are further extended to encompass very low-mass neutron stars, such as the central compact object in HESS J1731-347. To explore the influence of phase transitions on these relations, we construct an additional set of 100 hybrid equations of state, accounting for various features of the hadron--quark deconfinement transition. The macroscopic properties--such as masses, radii, and tidal deformabilities--of the resulting hybrid stars are found to be consistent with recent astrophysical observations. We further extend our analysis to establish quasi-universal relations for compact stars with more general core compositions, including nucleonic, heavy baryonic including entire baryon octet, and deconfined quark degrees of freedom. The possibility of the appearance of deconfined quark matter inside the core of low-mass neutron stars cannot be excluded from our EoS dataset. To this end, we derive relations among various macroscopic quantities using a comprehensive set consisting of 22 hadronic and 100 hybrid equations of state. Our results demonstrate that both the I-Love-Q relation for slowly rotating stars and the I-C-Q relation for rapidly rotating compact stars remain approximately universal. We observe that diverse core compositions degrade the quasi-universal behaviour, introducing variability of up to $\lesssim \mathcal{O}$(20\%). These results highlight the robustness and limitations of universal relations when extended to compact stars with diverse internal compositions and rotational profiles.
\end{abstract}
\maketitle

\section{Introduction}
\label{intro}
The structure and composition of neutron stars (NSs) are governed by the equation of state (EoS), which specifies the behaviour of matter at supranuclear densities and accounts for the possible emergence of exotic phases of matter \cite{Lattimer_2012_DkuaX, Lattimer_2016_DVZWU, _zel_2016_MkmxI, Oertel_2017_NDBAA, Baym_2018_fVTXR, Tolos_2020_aiFBV, Sedrakian_2023_UJvrR}. These compact stellar remnants, being extremely dense, provide a unique environment to probe such physics. The bulk properties of NSs such as mass, radius, and higher-order multipole moments are highly sensitive to the underlying EoS \cite{Lattimer_2001_WTceP, Ji_2019_aOrbX}. As these fundamental parameters influence the spacetime geometry surrounding these dense objects, precise observational constraints on these quantities can therefore yield vital insights into the state of matter under extreme conditions. For instance, observational data--especially mass and tidal deformability constraints inferred from gravitational-wave events like GW170817--enable tighter bounds on viable EoSs \cite{Bauswein_2017_JoXvH, De_2018_LcNIX, Malik_2018_qAfAT, Most_2018_JHlGX, Tews_2018_VRpfs}. Further, the advent of X-ray pulse profile modelling via NICER has significantly advanced the precision of pulsar property measurements, enabling estimations of mass, radius, and surface characteristics of NSs. Despite the limited understanding of matter beyond nuclear saturation density, such observational data provide critical constraints on the EoS \cite{Xie_2020_QhQne}. For example, Fonseca et al. \cite{Fonseca_2021_SOyMc} reported the mass of PSR J0740+6620 as 2.08$^{+0.07}_{-0.07} M\odot$ at a 68.3\% credibility level. This precise mass measurement of a heavy pulsar imposes significant limitations on the EoSs. However, the uncertainty in the corresponding radius still hampers definitive conclusions. More recent NICER results \cite{Miller_2019_dFNjM, Riley_2019_oQUYh, Miller_2021_pAcGM, Riley_2021_kbhlr, Salmi_2022_wcauZ} have yielded radius estimates with improved precision, thereby enhancing our ability to discriminate between competing EoS models. Moreover, PSR J1748-2446ad is reported to be a rapidly rotating pulsar with a spinning rate of 716 Hz, mass around 2 $M_\odot$, and radius within $\sim$ 16 km \cite{Hessels_2006_3QLXM}, puts further light on the composition of the interior of NSs. On the other hand, observations on the lower mass end, like the central compact object within HESS J1731-347, having mass M = 0.77$^{+0.20}_{-0.17} M\odot$ and radius R = 10.4$^{+0.86}_{-0.78}$ km \cite{Doroshenko_2022_xyimd} further complicates the EoS inferences.

From a nuclear theory standpoint, several EoS models that successfully reproduce nuclear saturation properties also remain compatible with current astrophysical observations \cite{Hebeler_2010_qcfiw, Oertel_2017_NDBAA, Dutra_2014_LtrKV, Tews_2018_cbmyx, Malik_2018_qAfAT}. Remarkably, for many such models, relationships between key stellar parameters--namely moment of inertia, quadrupole moment, and tidal deformability--exhibit a weak dependence on the specific microphysics of the EoS \cite{Yagi_2013_cPzvP, Maselli_2013_KEJBr}. These EoS-insensitive relations in NSs, known as universal relations, are especially useful for probing strong-field gravity. As such, universal relations among NS observables have attracted considerable attention in the literature \cite{Yagi_2013_cPzvP, Maselli_2013_KEJBr, Yagi_2013_yxUgU, Urbanec_2013_wNmiD, Baub_ck_2013_gJPHq, Doneva_2013_8E7jv, Doneva_2014_K2OBz, Cipolletta_2015_iJrxY, Tong_2025_rn, Roy_2025_ur}. The relation between the moment of inertia ($I$), tidal deformability ($\overline{\Lambda}$), and quadrupole moment ($Q$) is almost EoS independent for slowly rotating, non-magnetized neutron stars under small tidal deformation \cite{Yagi_2013_yxUgU, Urbanec_2013_wNmiD}. However, NSs possess surface magnetic fields ranging from $10^8$ to $10^{15}$ G, making them among the most magnetized objects known \cite{Phinney_1994_TmSXL, Manchester_2004_tPdWU, Thompson_1995_tjhSK, Kaspi_2004_nTcSP, Mereghetti_2008_NCHAj, Makishima_1999_DRwlJ}. To account for this, several studies have examined the persistence of universal relations under moderate magnetic fields \cite{Haskell_2013_cKUtF}. Moreover, tidal interactions significantly influence the gravitational-wave (GW) signal during NS binary inspirals \cite{Dudi_2018_jaQnY}, motivating beyond-adiabatic treatments \cite{Maselli_2012_WrTFq, Gagnon_Bischoff_2018_FCGMY, Castro_2021_xmbve} and extending the study of universal relations. Additionally, the observed pulsar spin periods, ranging from milliseconds to several seconds \cite{Manchester_2005_rjESW, Liu_2022_BdFHo}, support the relevance of extending these universal relations to account for rapid rotation \cite{Doneva_2013_8E7jv, Doneva_2014_K2OBz, Pappas_2014_TIiDy, Chakrabarti_2014_kBsdN, Sun_2020_QTRnD, Khosravi_Largani_2022_cvIJr}.

Many investigations have focused on nearly EoS-independent relations among macroscopic NS parameters \cite{Yagi_2013_cPzvP, Yagi_2013_yxUgU, Maselli_2013_KEJBr, Baub_ck_2013_gJPHq, Pappas_2014_TIiDy, Chakrabarti_2014_kBsdN, Chatterjee_2016_ijIxE, Marques_2017_dQf8a, Li_2018_yrEXW, Lenka_2019_VAoni, Raduta_2020_YKkHx, Li_2023_wnUwG}, deriving quasi-universal relations using a broad class of EoSs \cite{Urbanec_2013_wNmiD, Marques_2017_dQf8a, Raduta_2020_YKkHx, Suleiman_2021_lhDZ3, Khosravi_Largani_2022_cvIJr, Yang_2022_RbVSz, Saes_2022_KdfpG, Carlson_2023_0TLyw, Suleiman_2024_7cKmd}. These relations are also instrumental in testing general relativity and alternative gravity theories \cite{Pappas_2019_zqY5v, Sham_2014_EC3zI, Doneva_2016_QRTMQ}. Some works have employed generalized EoSs or phenomenological parameterizations within GR to obtain these relations \cite{Godzieba_2021_UBDLe, Saes_2022_KdfpG, Legred_2024_WRfSP, Suleiman_2024_7cKmd}. While much of this literature pertains to slowly rotating stars, extensions to rapid rotation have also been addressed \cite{Doneva_2013_8E7jv, Doneva_2014_K2OBz, Pappas_2014_TIiDy, Chakrabarti_2014_kBsdN, Breu_2016_WqIRD, Khosravi_Largani_2022_cvIJr}. A number of works also examine how heavy baryons modify NS structure and influence universal relations such as I-Love-Q and I-C-Q \cite{Bednarek_2012_dOXAn, Chatterjee_2016_ijIxE, Li_2018_yrEXW, Lenka_2019_VAoni, Li_2020_il2wz, Dexheimer_2021_Nc522, Logoteta_2021_MkalM, Sedrakian_2022_tiZFB, Marques_2017_dQf8a, Lenka_2019_VAoni, Raduta_2020_YKkHx, Li_2023_wnUwG}. Incorporating heavy baryonic components in the EoS models broadens the accessible region in the pressure-energy density plane. Particularly, heavy baryonic degrees of freedom, such as hyperons and $\Delta$ resonances, tend to soften the pressure-density relation, thereby reducing the maximum mass, altering radii, and affecting tidal deformabilities of NSs \cite{Weber_2005_SYDRw, Tolos_2016_kVYCR}. We construct a dataset of 22 hadronic EoSs that include the full baryonic octet along with nucleonic degrees of freedom. Quantitative analyses reveal that heavy baryonic matter typically lowers the maximum mass while increasing the tidal deformability due to reduced stiffness \cite{Bednarek_2012_dOXAn, Logoteta_2021_MkalM}, thereby affecting the quasi-universal relations.

Furthermore, the hadron-quark phase transition plays a vital role in shaping the high-density behaviour of the EoS, influencing not just the mass and radius but also core-collapse dynamics and post-merger GW signals \cite{Burgio_2002_gVOwJ, Hanauske_2019_xvUXm, Jakobus_2022_rTUic}. Numerous papers address the role of a hadron-quark phase transition in the context of recent astrophysical observations \cite{Bonanno_2012_Ca0lZ, Ayvazyan_2013_NZ1QJ, Pereira_2018_LsPLH, Nandi_2018_YZU8D, Li_2021_e1GSh}. In addition, several authors have explored EoS insensitive relations for quark stars (QSs) and hybrid stars (HSs) \cite{Urbanec_2013_wNmiD, Yagi_2013_cPzvP, Yagi_2014_lZhJS, Yagi_2017_uWwLw, Bozzola_2019_861n9, Yeung_2021_dc0JE, Khosravi_Largani_2022_cvIJr, Roy_2025_ur}. Refs. \cite{Urbanec_2013_wNmiD, Yeung_2021_dc0JE, Burikham_2022_Kq9sd} employ slow-rotation approximation to investigate the quasi-universal relations between I, deformation, Q for NSs, QSs and HSs. These relations are further extended for rapidly rotating QSs and HSs, with a set of limited EoSs \cite{Yagi_2013_cPzvP, Yagi_2014_lZhJS, Yagi_2017_uWwLw, Bozzola_2019_861n9, Roy_2025_ur}. A few subsequent attempts are also made to establish universal relations involving macroscopic quantities like maximum mass, moment of inertia, compactness, etc., of rapidly rotating HSs with a broader set of EoSs \cite{Khosravi_Largani_2022_cvIJr, Papigkiotis_2023_a8kst}. Despite these efforts, a systematic and comprehensive exploration of universal relations that incorporates both full baryon octet and hadron-quark transitions under arbitrary rotation is still lacking. To address this, we construct an additional set of 100 hybrid EoSs that incorporate nucleonic and heavy baryonic matter and allow for transitions to deconfined quark matter at sufficiently high densities. We then derive quasi-universal relations using both the hadronic and hybrid EoS sets--linking moment of inertia, compactness, and quadrupole moment--for rapidly rotating NSs and HSs, collectively referred to as compact stars (CSs). A detailed comparison with earlier results elucidates how core composition influences these approximately universal relations. In sections \ref{eos_model} and \ref{equilibrium_model}, we review the EoS models employed in this work and the equations governing the equilibrium configurations for both slowly and rapidly rotating CSs. Our findings and their comparison with prior studies are discussed in section \ref{result_discussion}, followed by conclusions in section \ref{conclusion}.

\begin{figure}
  \centering
  \captionsetup{justification=raggedright}
  \subfloat[Pressure vs energy density for Hadronic EoSs]{\includegraphics[width=1.0\linewidth]{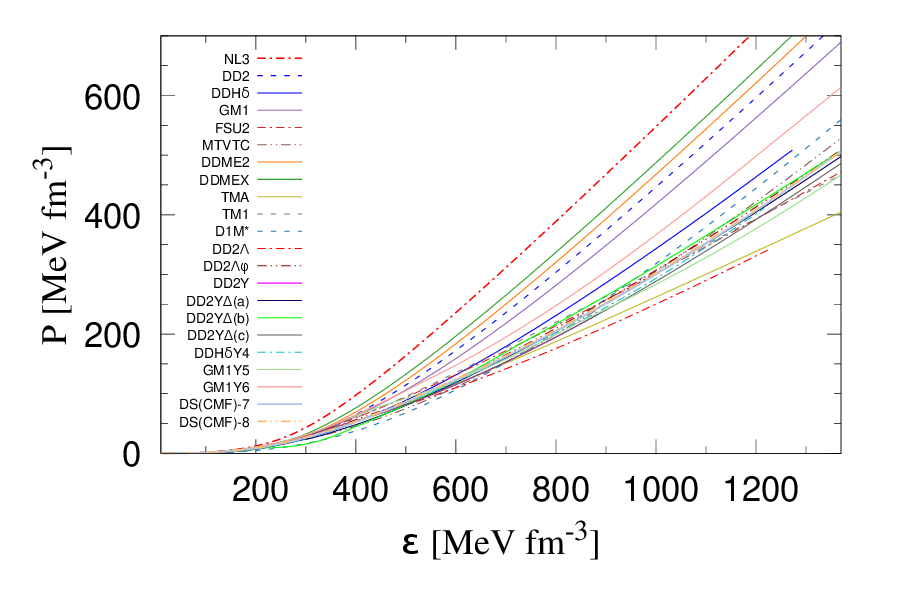}}\hfill
  \subfloat[$c_s^2$ vs energy density for Hadronic EoSs.]{\includegraphics[width=1.0\linewidth]{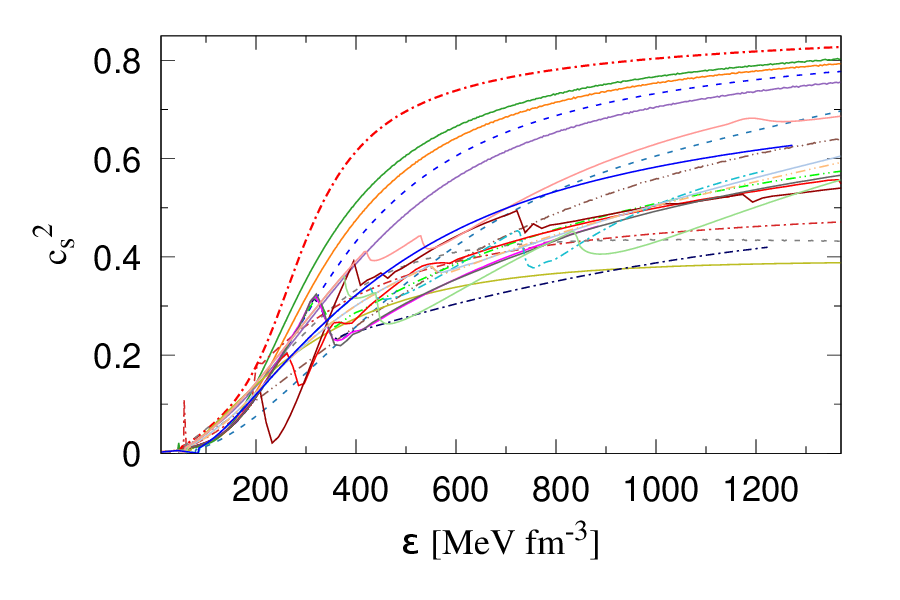}}
\caption{Hadronic EoSs for cold NS matter incorporating heavy baryon degrees of freedom. Subfigure (a) shows the pressure–energy density relations for various EoS models, while subfigure (b) presents the corresponding speed of sound squared, $c_s^2$, profiles. The colour coding is consistent across both subfigures.}
\label{fig1H_eos} 
\end{figure}

\begin{figure}
  \centering
  \captionsetup{justification=raggedright}
  \subfloat[Pressure vs energy density for the full set of EoSs.]{\includegraphics[width=1.0\linewidth]{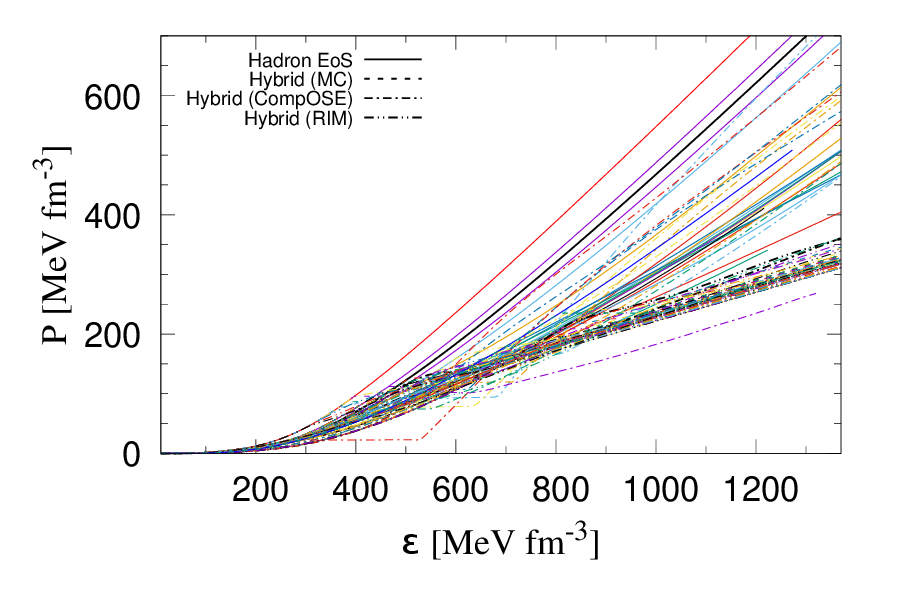}}\hfill
  \subfloat[$c_s^2$ vs energy density for full set of EoSs.]{\includegraphics[width=1.0\linewidth]{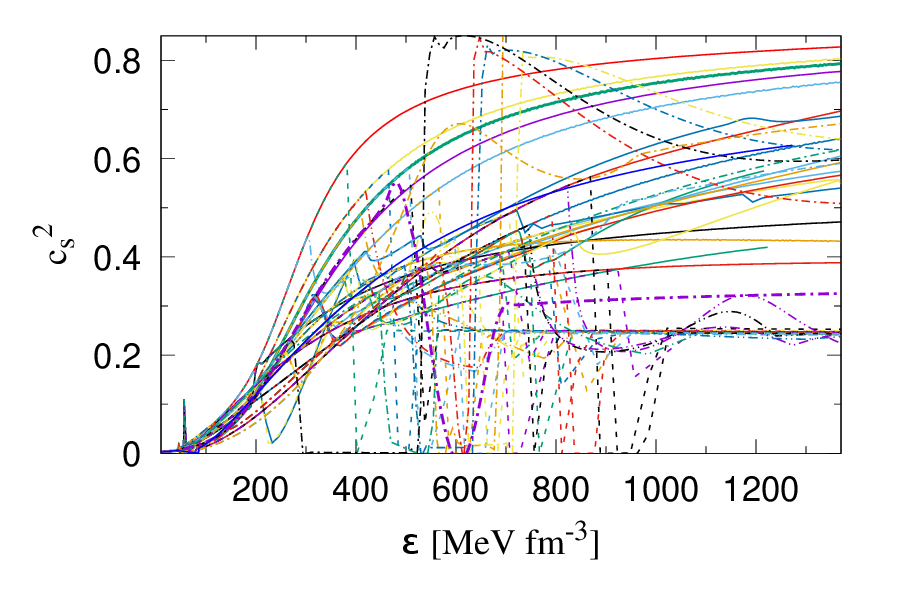}}
\caption{Hybrid EoSs, along with the hadronic EoSs, are shown in subfigure (a). Solid curves represent the hadronic EoSs, while dashed curves correspond to hybrid EoSs constructed via the Maxwell phase transition. Single-dot dashed curves indicate hybrid EoSs obtained from CompOSE, and double-dot dashed curves represent those generated using the replacement interpolation method (RIM). The associated speed of sound squared, $c_s^2$, for the considered EoSs is presented in subfigure (b). The colour scheme is the same across both subfigures.}
\label{fig1A_eos} 
\end{figure}


\begin{figure}
  \centering
  \captionsetup{justification=raggedright}
  \subfloat[Static limit of M-R]{\includegraphics[width=1.0\linewidth]{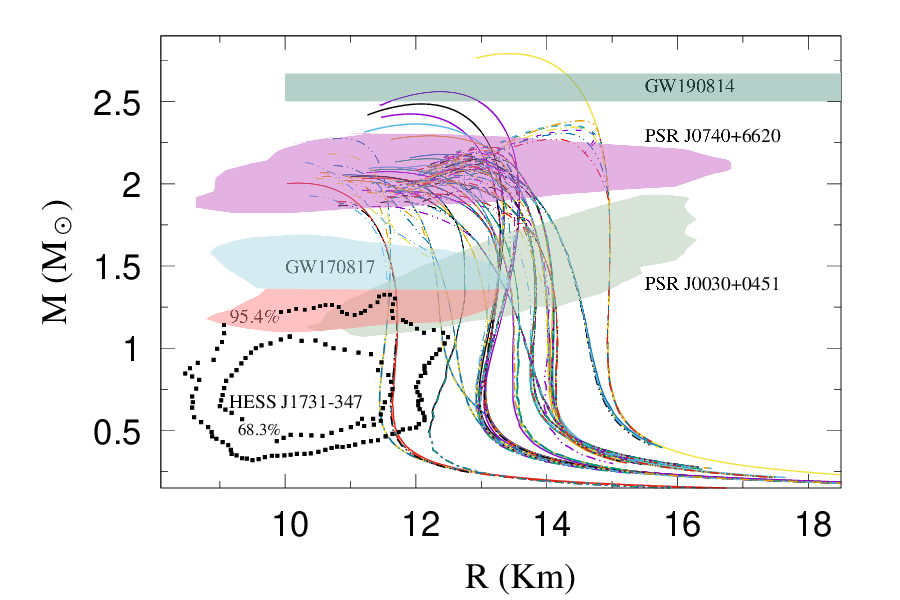}}\hfill
  \subfloat[Keplerian limit of M-R]{\includegraphics[width=1.0\linewidth]{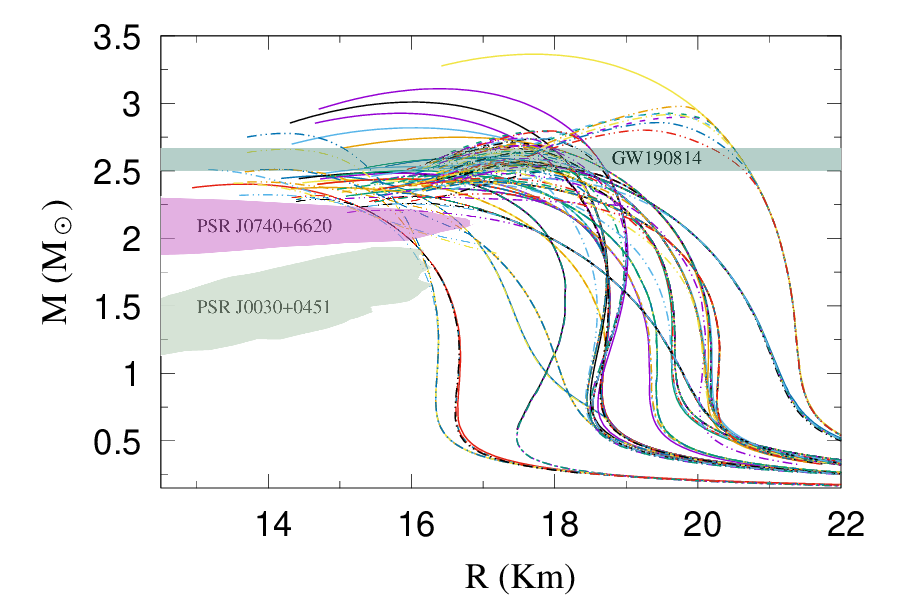}}
  \caption{ Mass-radius relations of CSs are shown in subfigures (a) and (b) for the static and Keplerian limits, respectively, using the full set of EoSs. NICER measurements for PSR J0030+0451 \cite{Miller_2019_dFNjM, Riley_2019_oQUYh} and PSR J0740+6620 \cite{Miller_2021_pAcGM, Riley_2021_kbhlr} (95\% credible intervals) are overlaid. Subfigure (a) also includes the NS radius constraints inferred from GW170817 \cite{Abbott_2018_EX9Nk}, NS mass-radius constraints for central compact object in HESS J1731--347 \cite{Doroshenko_2022_xyimd} (68.3\% and 95.4\% credibility intervals), and the mass range estimated for the secondary component of GW190814 \cite{Abbott_2020_jIBsD} is indicated. Identical colour codes are used in both subfigures.}
  \label{fig2_M-R}
\end{figure}

\begin{figure}
  \centering
  \captionsetup{justification=raggedright}
  {\includegraphics[width=1.0\linewidth]{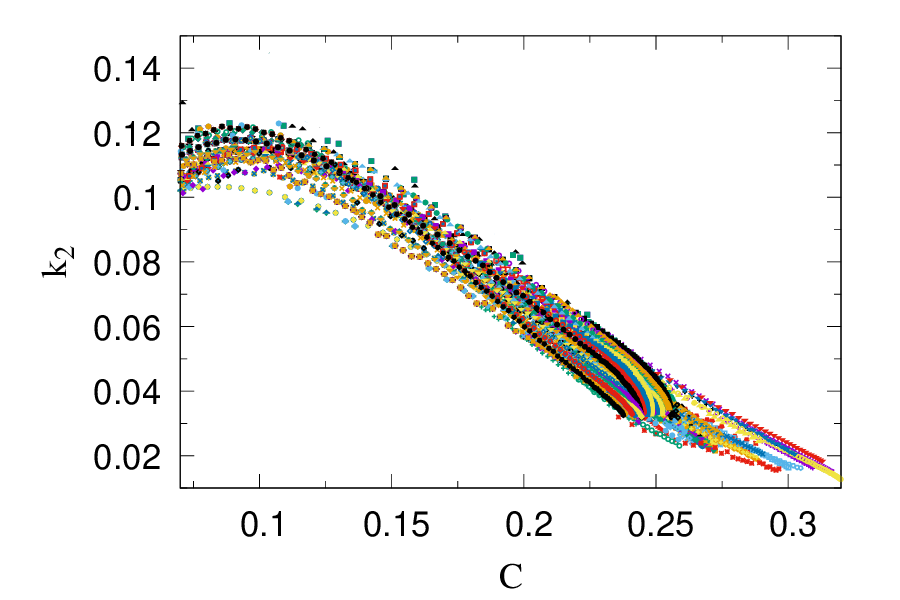}}
  \caption{Variation of the dimensionless tidal Love number $k_2$ with compactness $C$ for the enitre EoS dataset.}
  \label{fig3_k2-C}
\end{figure}

\begin{figure}
  \centering
  \captionsetup{justification=raggedright}

  \includegraphics[width=0.8\linewidth]{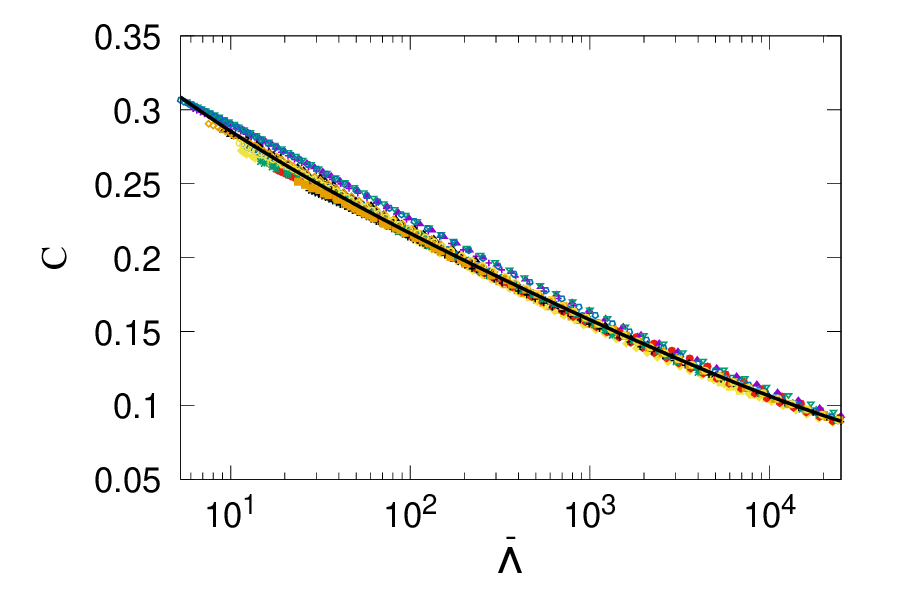}
  \vspace{-0.25cm}
  \subfloat[Results for NSs]{\includegraphics[width=0.8\linewidth]{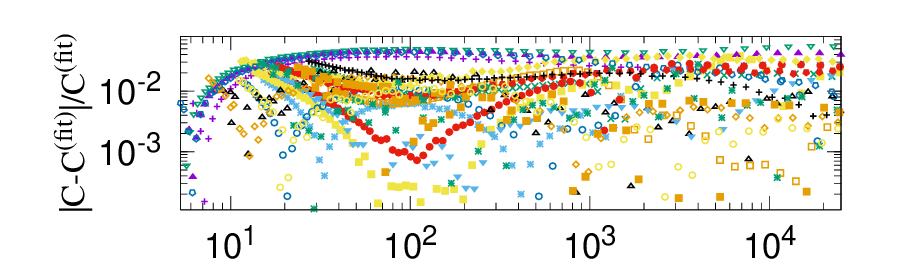}}

  \vspace{0.5cm} 

  \includegraphics[width=0.8\linewidth]{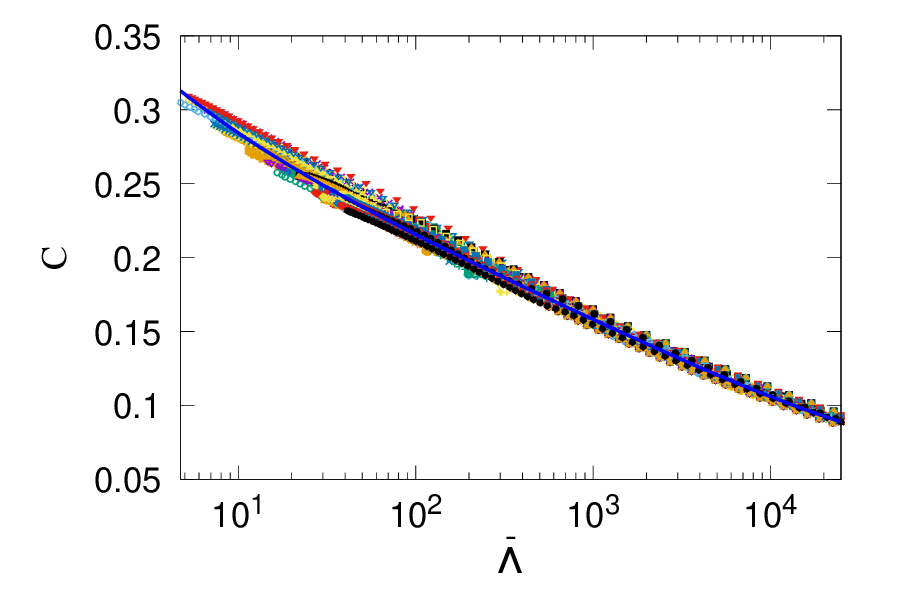}
  \vspace{-0.25cm}
  \subfloat[Results for NSs+HSs]{\includegraphics[width=0.8\linewidth]{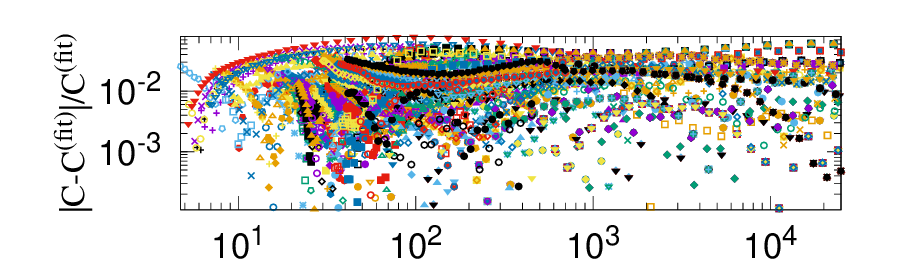}}

  \caption{$C$–$\overline{\Lambda}$ relations obtained using (a) hadronic EoSs and (b) both hadronic and hybrid EoSs. The top panels display the data and fit; the bottom panels show corresponding fractional deviations.}
  \label{fig4_C-L}
\end{figure}

\begin{figure}
  \centering
  \captionsetup{justification=raggedright}
  
  \includegraphics[width=0.8\linewidth]{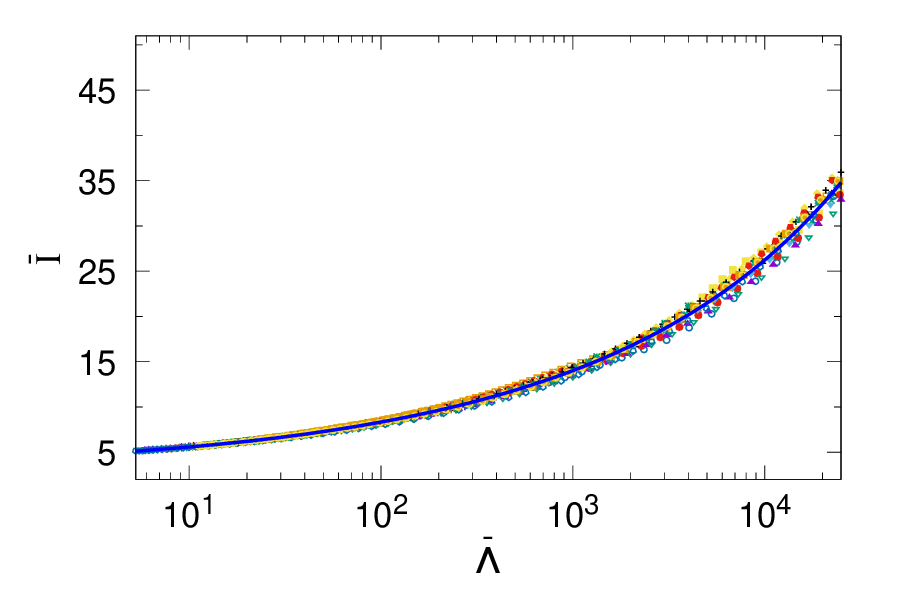}
  \vspace{-0.25cm}
  \subfloat[Results for NSs]{\includegraphics[width=0.8\linewidth]{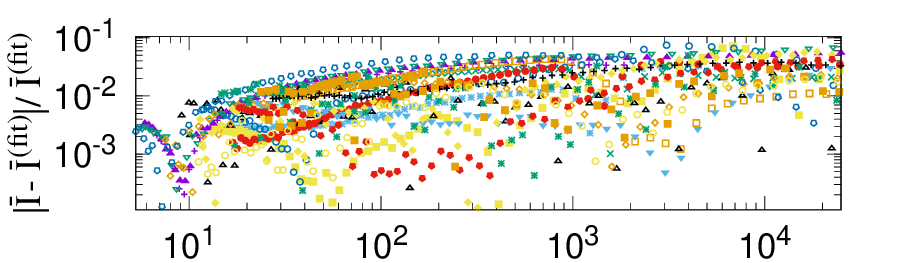}}

  \vspace{0.5cm} 

  \includegraphics[width=0.8\linewidth]{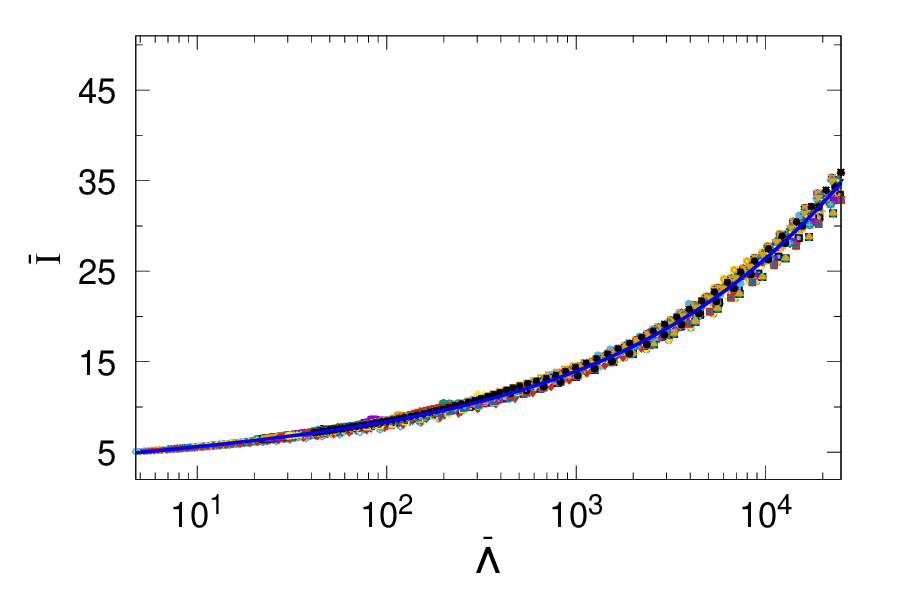}
  \vspace{-0.25cm}
  \subfloat[Results for NSs+HSs]{\includegraphics[width=0.8\linewidth]{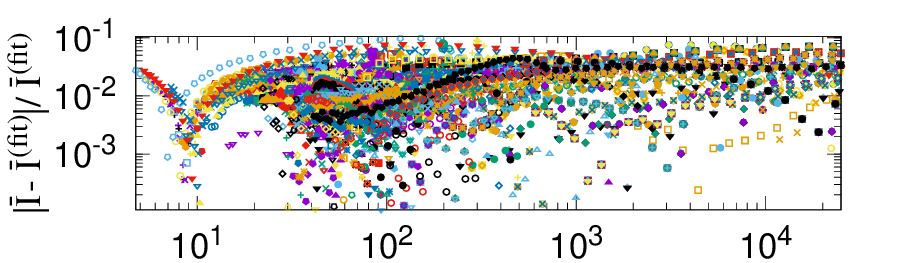}}
  
  \caption{The $\overline{I}$–$\overline{\Lambda}$ relation for slowly rotating (a) NSs and (b) combined NSs and HSs. The upper panels show the calculated data points and corresponding fits across all EoSs, while the lower panels display the fractional deviations between the fits and numerical results.}
  \label{fig5_Ib-L}
\end{figure}

\begin{figure}
  \centering
  \captionsetup{justification=raggedright}
  
  \includegraphics[width=0.8\linewidth]{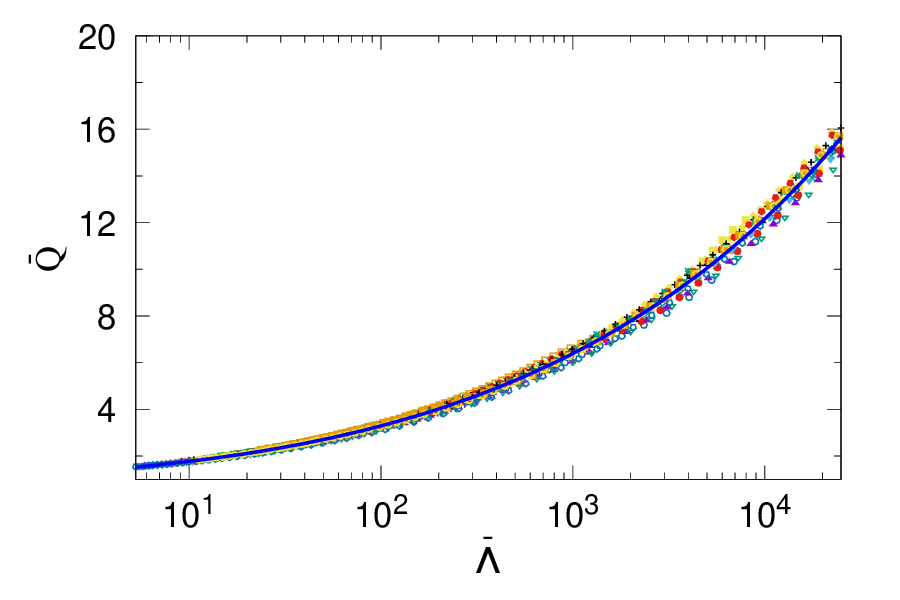}
  \vspace{-0.25cm}
  \subfloat[Results for NSs]{\includegraphics[width=0.8\linewidth]{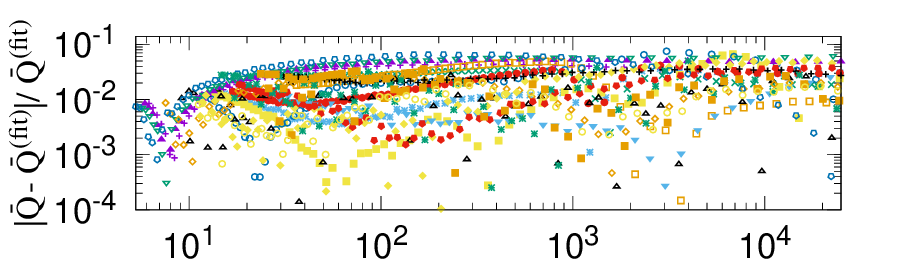}}

  \vspace{0.5cm} 

  \includegraphics[width=0.8\linewidth]{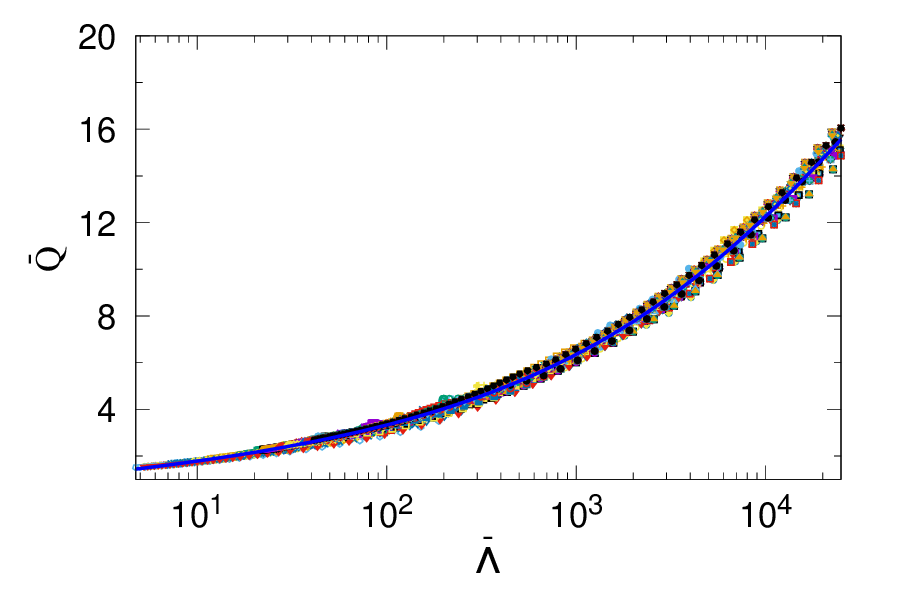}
  \vspace{-0.25cm}
  \subfloat[Results for NSs+HSs]{\includegraphics[width=0.8\linewidth]{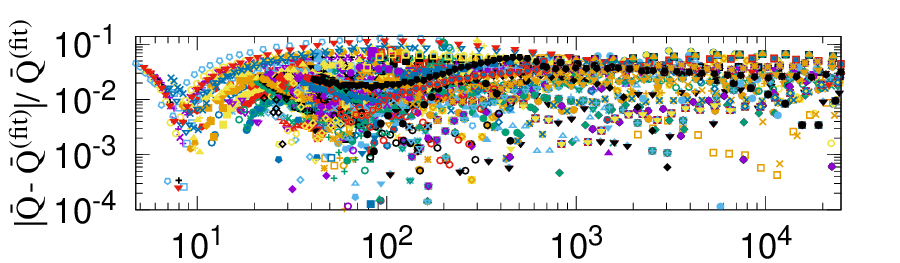}}
  
  \caption{The $\overline{Q}$–$\overline{\Lambda}$ relations for slowly rotating (a) NSs and (b) combined NSs and HSs. The upper panels show the calculated relations, while the lower panels display the fractional deviations.}
  \label{fig7_Qb-L}
\end{figure}

\begin{figure}
  \centering
  \captionsetup{justification=raggedright}

  \includegraphics[width=0.8\linewidth]{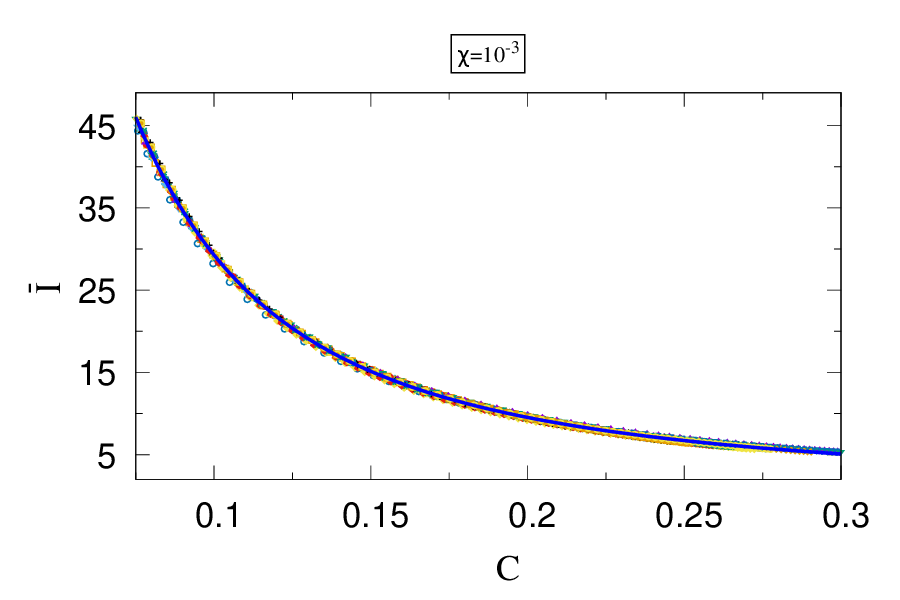}
  \vspace{-0.25cm}
  \subfloat[Results for NSs]{\includegraphics[width=0.8\linewidth]{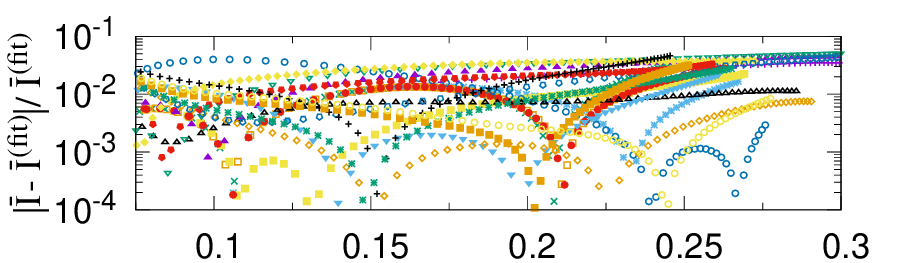}}

  \vspace{0.5cm} 

  \includegraphics[width=0.8\linewidth]{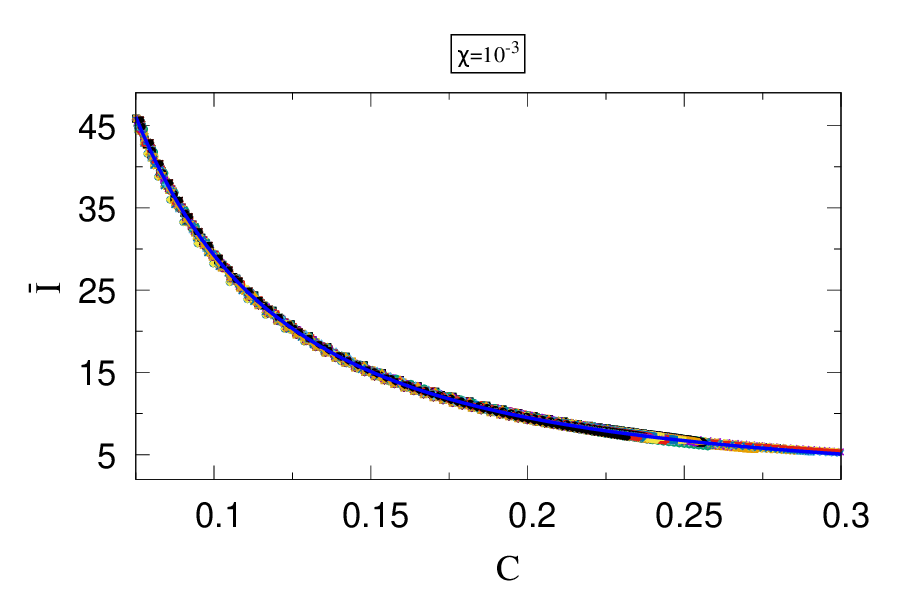}
  \vspace{-0.25cm}
  \subfloat[Results for NSs+HSs]{\includegraphics[width=0.8\linewidth]{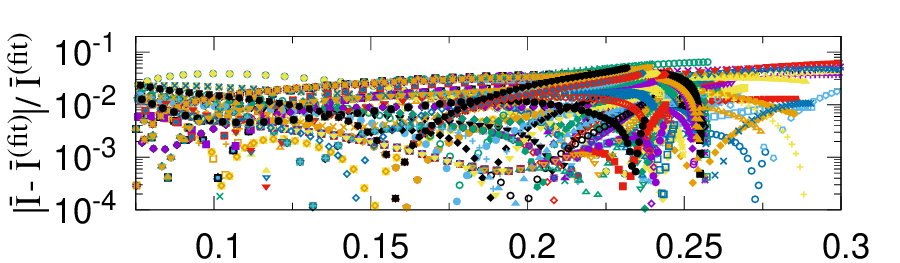}}

  \caption{Slowly rotating $\overline{I}$–$C$ relations at $\chi = 10^{-3}$ for (a) NSs and (b) NSs + HSs. Top panels show the fitted relations; bottom panels present the relative deviations.}
  \label{fig8A_I-C_sr}
\end{figure}

\begin{figure}
  \centering
  \captionsetup{justification=raggedright}
  
  \includegraphics[width=0.8\linewidth]{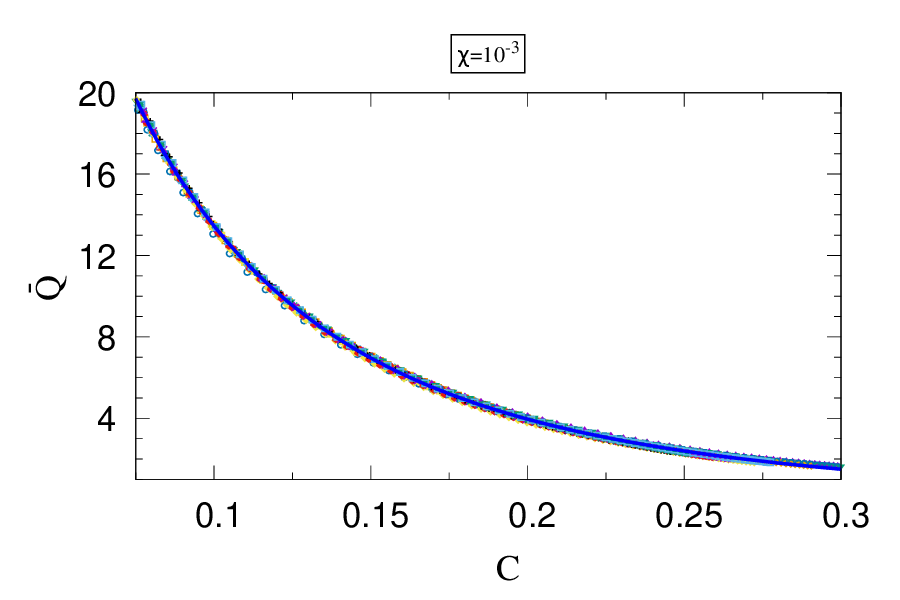}
  \vspace{-0.25cm}
  \subfloat[Results for NSs]{\includegraphics[width=0.8\linewidth]{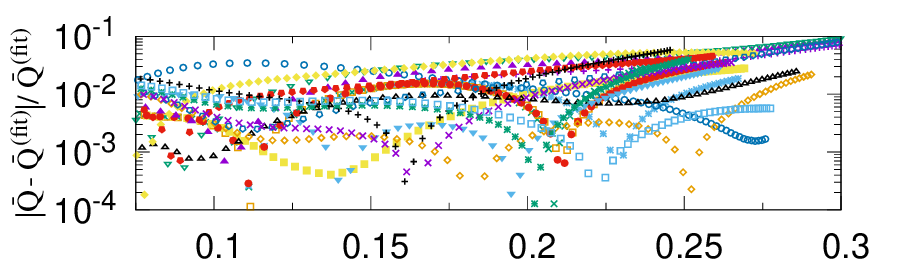}}

  \vspace{0.5cm} 

  \includegraphics[width=0.8\linewidth]{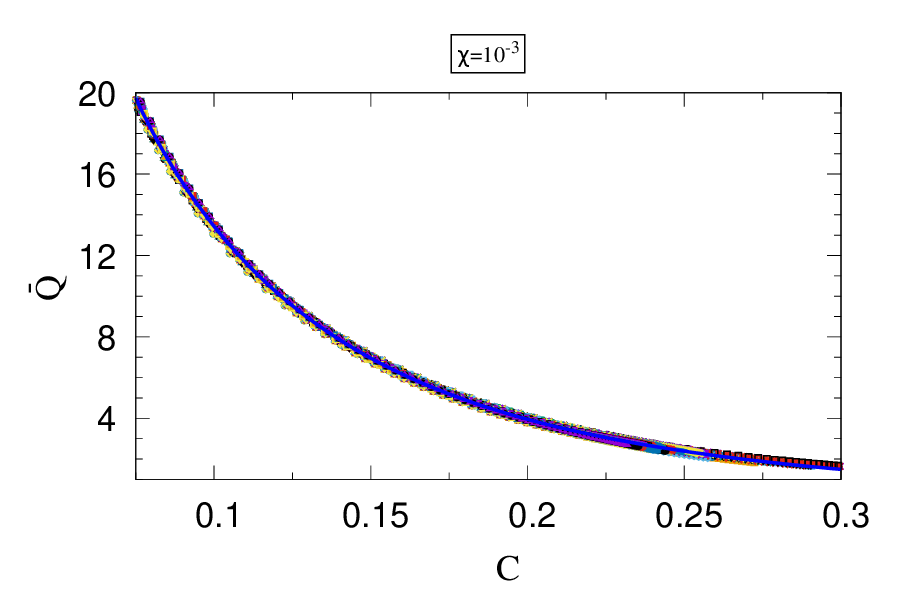}
  \vspace{-0.25cm}
  \subfloat[Results for NSs+HSs]{\includegraphics[width=0.8\linewidth]{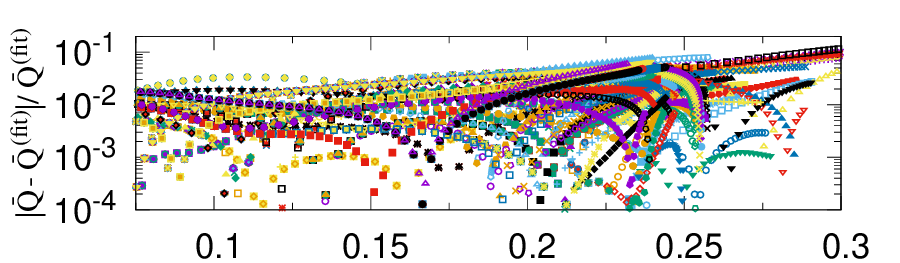}}

  \caption{$\overline{Q}$–$C$ relations for slowly rotating (a) NSs and (b) NSs + HSs at $\chi = 10^{-3}$. The upper panels show the computed data with fits, and the lower panels present the fractional deviations.}
\label{fig9A_Q-C_sr}
\end{figure}

\begin{figure}
  \centering
  \captionsetup{justification=raggedright}

  \includegraphics[width=0.8\linewidth]{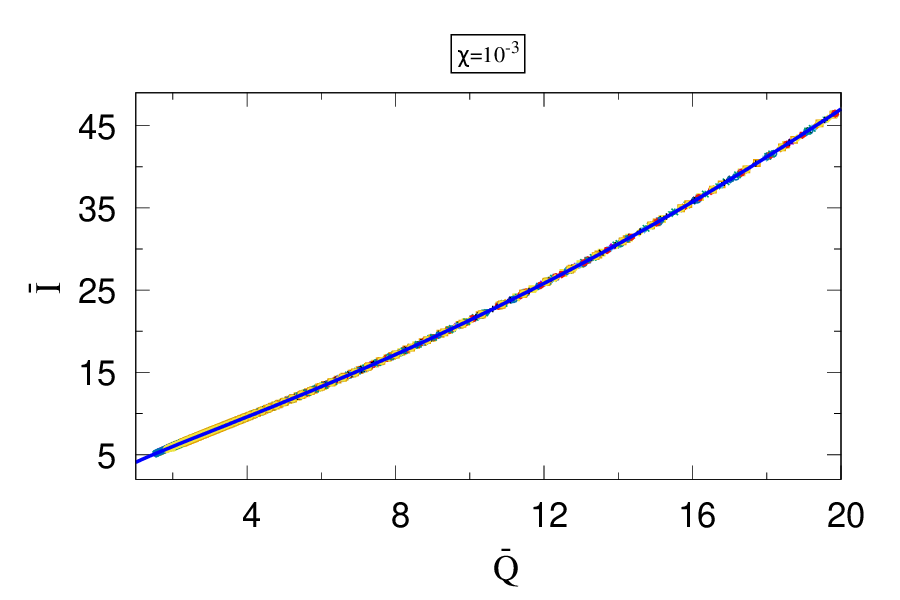}
  \vspace{-0.25cm}
  \subfloat[Results for NSs]{\includegraphics[width=0.8\linewidth]{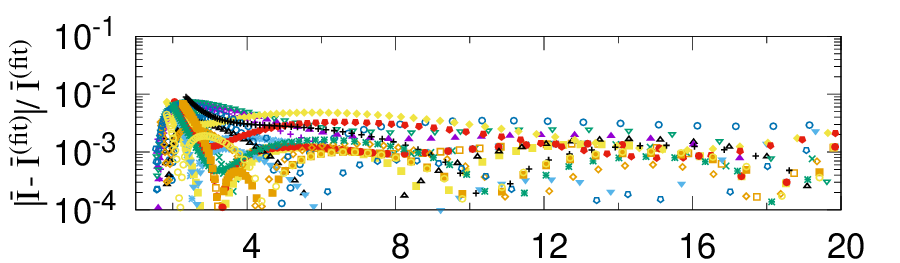}}

  \vspace{0.5cm} 

  \includegraphics[width=0.8\linewidth]{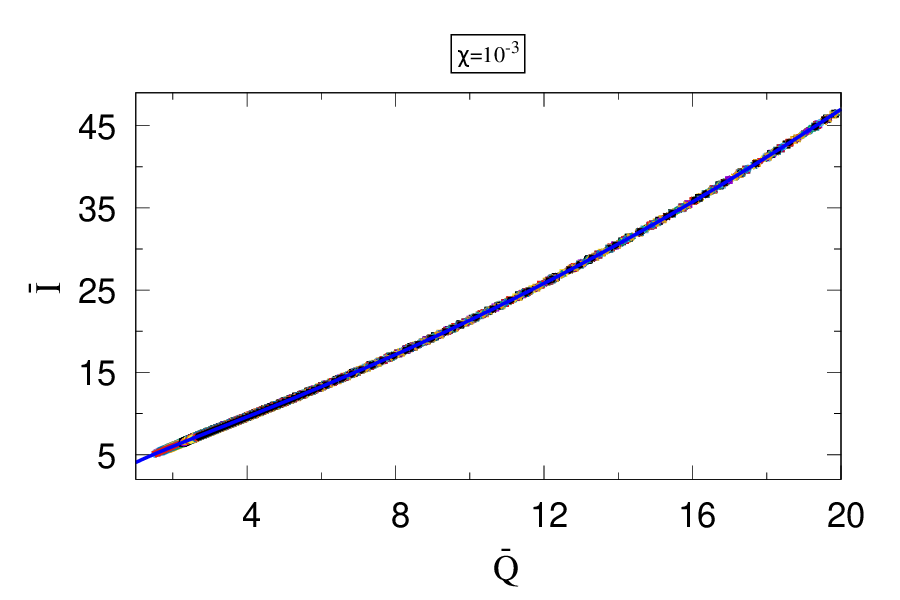}
  \vspace{-0.25cm}
  \subfloat[Results for NSs+HSs]{\includegraphics[width=0.8\linewidth]{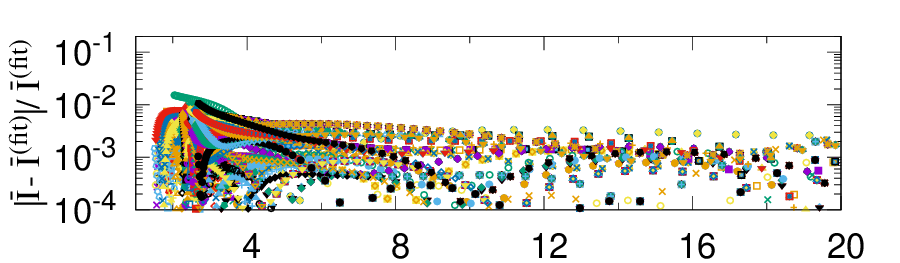}}

  \caption{$\overline{I}$–$\overline{Q}$ relations for slowly rotating configurations at $\chi = 10^{-3}$ are shown for (a) NSs and (b) NSs + HSs. The top panels display the computed data along with the fitted curves, while the bottom panels present the corresponding relative deviations.}
  \label{fig6_I-Q_sr}
\end{figure}
\begin{figure}
  \centering
  \captionsetup{justification=raggedright}

  \includegraphics[width=0.8\linewidth]{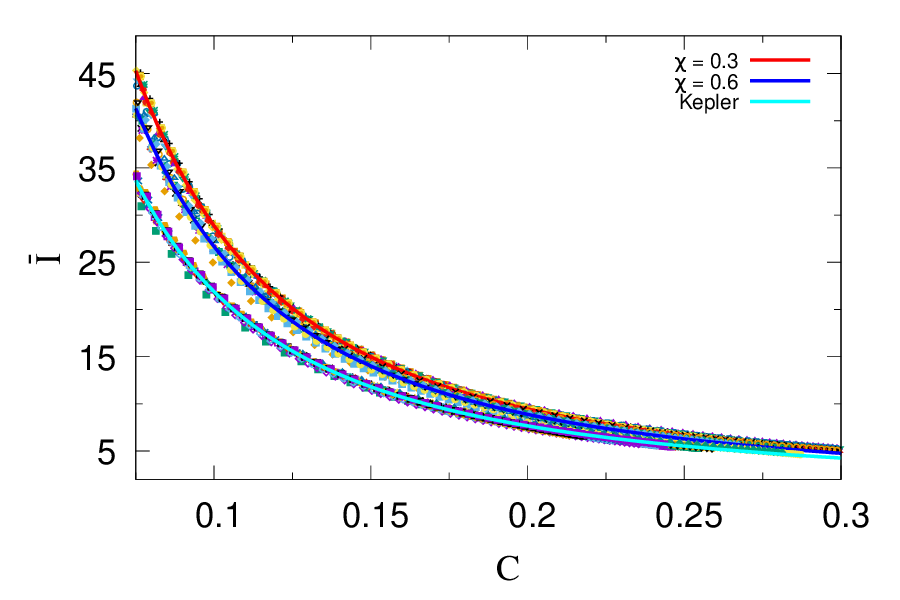}
  \vspace{-0.25cm}
  \subfloat[Results for NSs]{\includegraphics[width=0.8\linewidth]{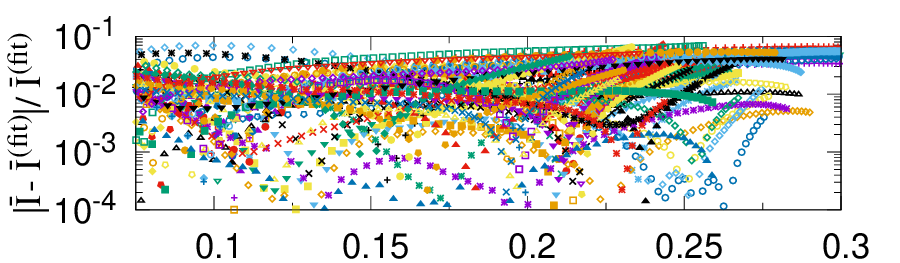}}

  \vspace{0.5cm} 

  \includegraphics[width=0.8\linewidth]{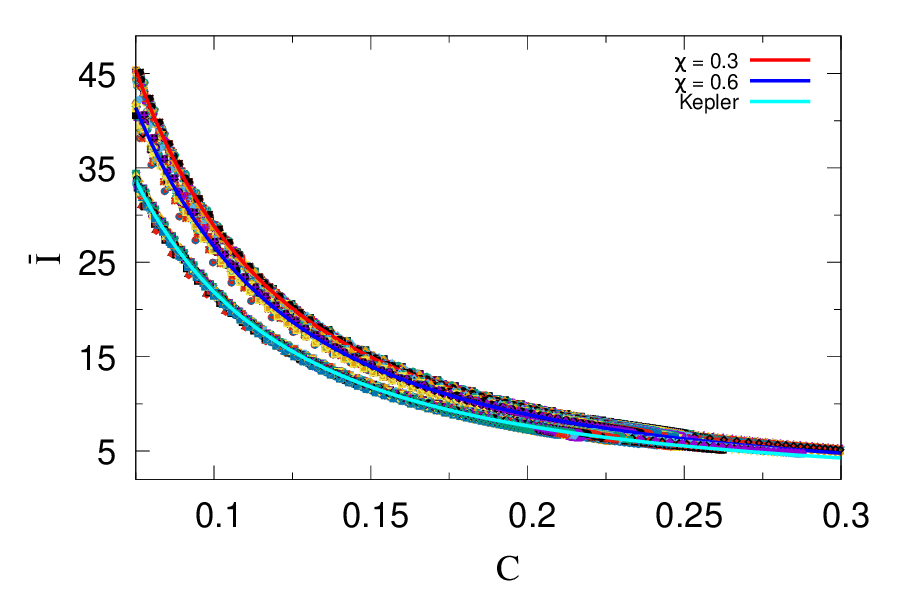}
  \vspace{-0.25cm}
  \subfloat[Results for NSs+HSs]{\includegraphics[width=0.8\linewidth]{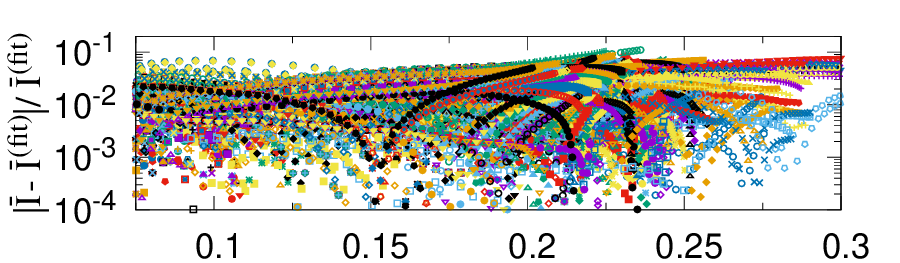}}

  \caption{$\overline{I}$–$C$ relations for rapidly rotating configurations are shown for (a) NSs and (b) NSs + HSs, corresponding to $\chi = 0.3$, $\chi = 0.6$, and the maximally rotating (Kepler) sequences. The upper panels display the computed data with fitted curves, while the lower panels show the relative deviations.}
  \label{fig8B_I-C}
\end{figure}
\begin{figure}
  \centering
  \captionsetup{justification=raggedright}

  \includegraphics[width=0.8\linewidth]{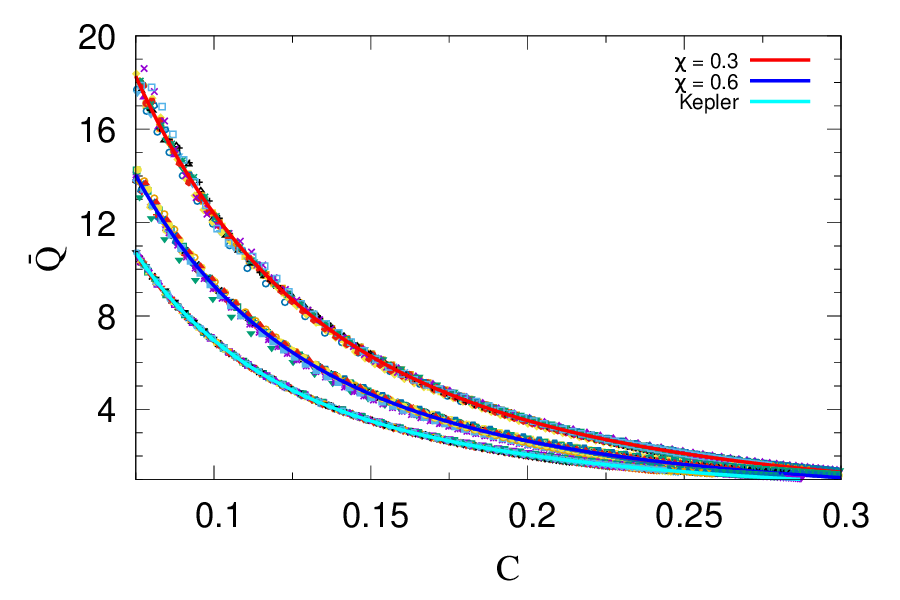}
  \vspace{-0.25cm}
  \subfloat[Results for NSs]{\includegraphics[width=0.8\linewidth]{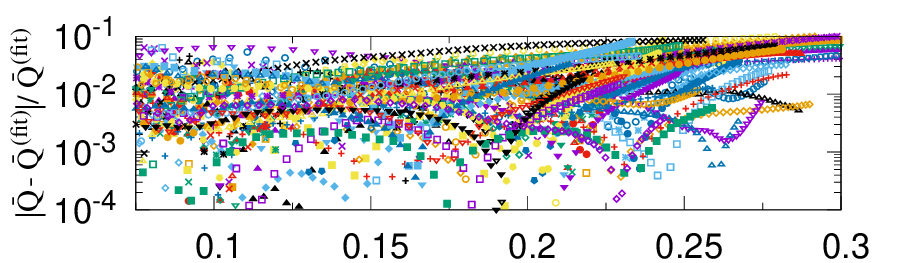}}

  \vspace{0.5cm} 

  \includegraphics[width=0.8\linewidth]{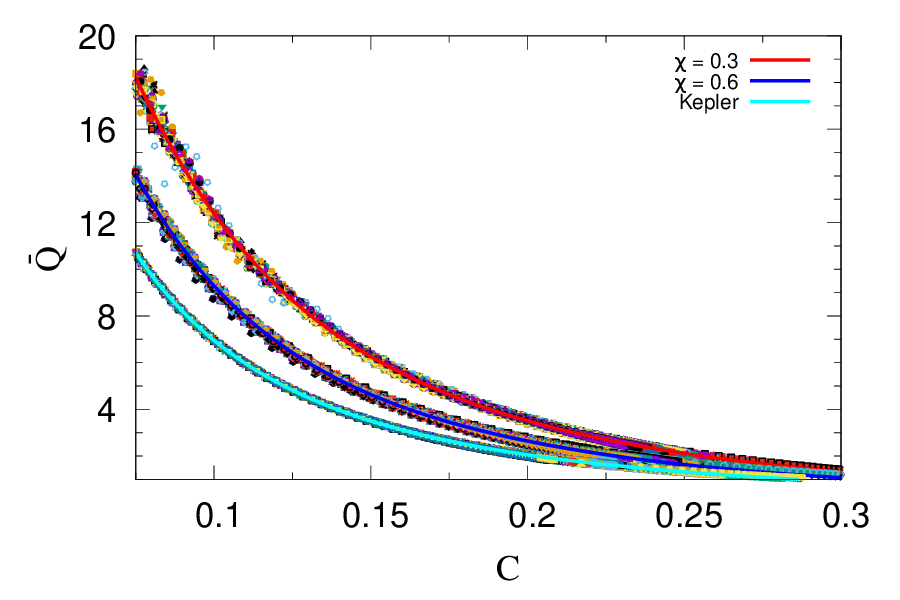}
  \vspace{-0.25cm}
  \subfloat[Results for NSs+HSs]{\includegraphics[width=0.8\linewidth]{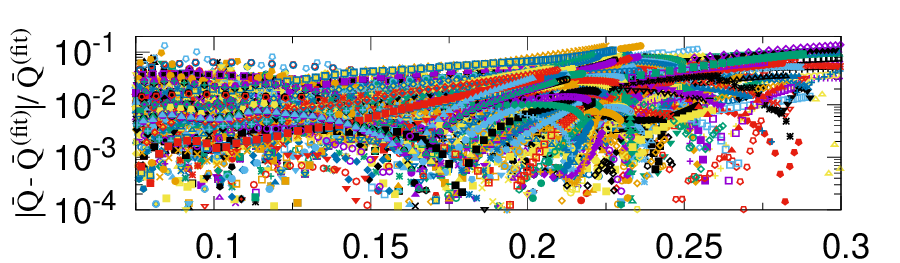}}

  \caption{$\overline{Q}$–$C$ relations for rapidly rotating configurations are presented for (a) NSs and (b) NSs + HSs, corresponding to $\chi = 0.3$, $\chi = 0.6$, and the maximally rotating (Kepler) sequences. The upper panels display the numerical results along with the fitted curves, while the lower panels show the corresponding relative deviations.}
  \label{fig9B_Q-C}
\end{figure}
\begin{figure}
  \centering
  \captionsetup{justification=raggedright}

  \includegraphics[width=0.8\linewidth]{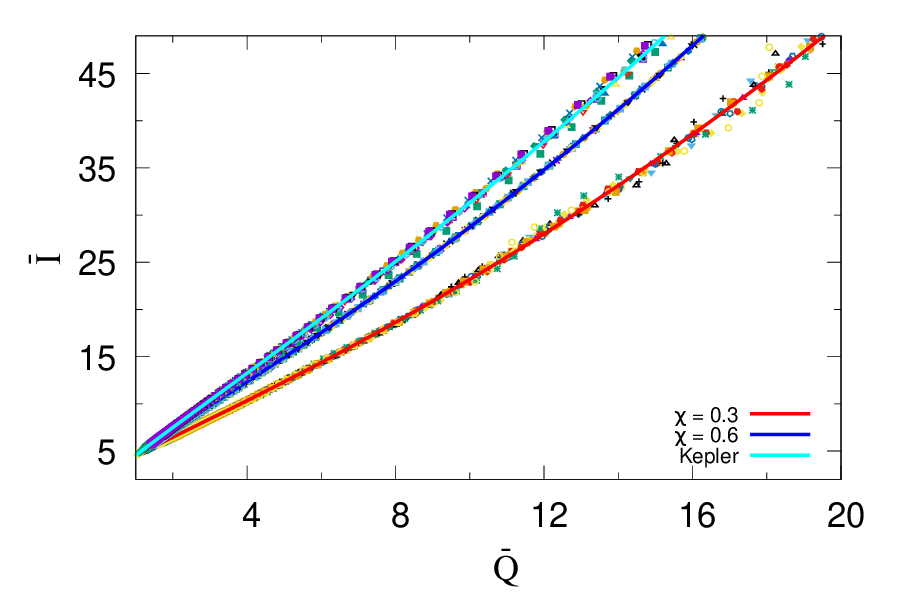}
  \vspace{-0.25cm}
  \subfloat[Results for NSs]{\includegraphics[width=0.8\linewidth]{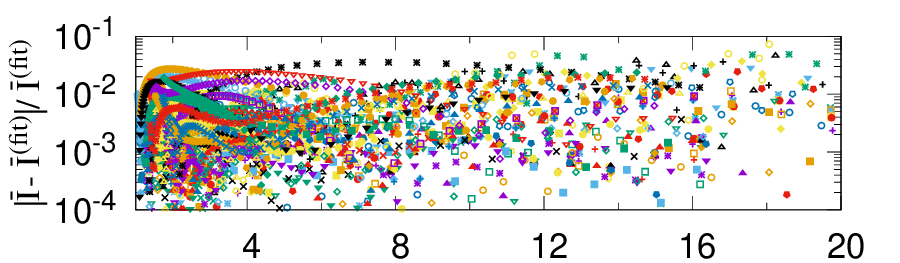}}

  \vspace{0.5cm} 

  \includegraphics[width=0.8\linewidth]{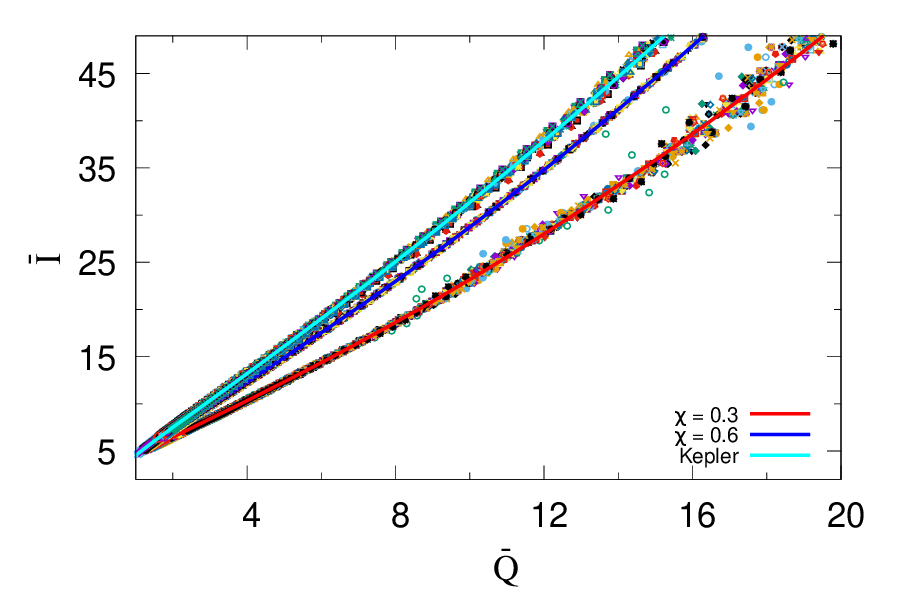}
  \vspace{-0.25cm}
  \subfloat[Results for NSs+HSs]{\includegraphics[width=0.8\linewidth]{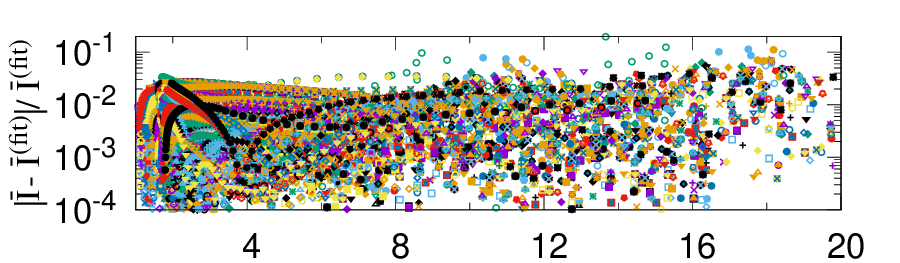}}
  \caption{$\overline{I}$–$\overline{Q}$ relations for rapidly rotating configurations are shown for (a) NSs and (b) NSs + HSs, corresponding to $\chi = 0.3$, $\chi = 0.6$, and the maximally rotating (Kepler) sequences. The upper panels present the numerical results with the fitted curves, and the lower panels display the relative deviations.}
  \label{fig10B_I-Q}
\end{figure}

\section{Equation of State}
\label{eos_model}
In this study, we adopt a range of hadronic EoSs constructed within the framework of the Relativistic Mean Field (RMF) model \cite{Typel_2010_CZbbi}. These include the NL3 parametrization \cite{Lalazissis_1997_NL3}, which yields a very stiff EoS, and the DD2 EoS \cite{Hempel_2010_mRz7r} for interacting nucleons at zero temperature. The DDH$\delta$ EoS \cite{Gaitanos_2004_DDHdelta} is employed to account for the effects of the scalar isovector $\delta$-meson field. Additional EoS variations are introduced via the DDME2, DDMEX, GM1, TM1, and MTVTC EoSs \cite{Xia_2022_DDME2, Xia_2022_DDMEX}, developed using the Thomas-Fermi approximation. We also include the FSU2 EoS \cite{Shen_2011_FSU2} and the TMA EoS \cite{Toki_1995_TMA, Hempel_2010_mRz7r} for $\beta$-equilibrated cold neutron star matter. Furthermore, the D1M* EoS \cite{Vi_as_2021_D1M*}, based on non-relativistic density functional theory with effective Gogny interactions, is incorporated into our hadronic EoS set. All EoS tables used in this work are obtained from the CompOSE database \cite{obspmHome, doiCompOSEReference, Oertel_2017_NDBAA, Typel_2015_gH0QU}.

\subsection{Nuclear Matter with Heavy Baryons}
\label{eos_nuclear}
We consider the possibility that heavy baryon degrees of freedom begin to appear in NS matter at around 2–3 times the nuclear saturation density \cite{Balberg_1999_XcODI, Drago_2014_kXwYr}. These exotic components play a significant role in shaping the integral properties of NSs \cite{Balberg_1999_XcODI, CHAMEL_2013_ma6UJ}. Accordingly, we adopt several EoSs that explicitly account for such phases. In particular, we employ the DD2$\Lambda$ and DD2$\Lambda\phi$ EoSs from Ref. \cite{Banik_2014_QOKgS}. The attractive potential depth for $\Lambda$ hyperons in the DD2$\Lambda$ EoS reduces the maximum mass of the static NS sequence relative to that predicted by the DD2 EoS. In contrast, the repulsive $\Lambda$–$\Lambda$ interaction mediated by the strange $\phi$ meson in the DD2$\Lambda\phi$ EoS slightly increases the maximum mass. Unless stated otherwise, the term "maximum mass" refers to the static mass of the last stable configuration.

To incorporate the effects of the full baryonic octet, we employ the DD2Y EoS from Ref. \cite{Marques_2017_dQf8a}, which includes $\sigma$-meson-mediated hyperon interactions. We also consider the DDH$\delta$Y4, GM1Y5, and GM1Y6 EoSs \cite{Oertel_2015_DDHDY4}, which describe cold neutron star matter in $\beta$-equilibrium including all members of the baryon octet. In addition, we adopt three zero temperature EoSs—DD2Y$\Delta$(a), DD2Y$\Delta$(b), and DD2Y$\Delta$(c)—from Ref. \cite{Raduta_2022_mFVlJ}, which model $\Delta$-admixed nuclear matter, each corresponding to a different $\Delta$ potential depth. We further include two EoSs based on the chiral mean-field (CMF) model, DS(CMF)-7 and DS(CMF)-8 \cite{obspmHome}, taken from Ref. \cite{Dexheimer_2008_FEd1R}, which describe $\beta$-equilibrated cold NS matter with hyperons and other resonance states.

It is worth noting that the symmetry energy and its slope for the EoSs introduced above are consistent with the constraints drawn from PREX-II neutron skin thickness measurements \cite{Reinhard_2021_yDCxi, Essick_2021_AG8v4}. However, the EoSs predict larger values of the symmetry energy slope parameter than those favoured by the CREX measurements \cite{Miyatsu_2023_XrwEA}. Moreover, the selected set of EoSs not only places the nuclear incompressibility parameter within the acceptable range but also supports negative symmetry incompressibilities, as preferred by recent astrophysical constraints \cite{Essick_2021_AG8v4, Essick_2021_fCXCe}. The mass–radius constraints from NICER observations, which generally suggest the need for stiffer EoSs \cite{Cromartie_2019_NtRiY, Miller_2021_pAcGM, Riley_2019_oQUYh, Riley_2021_kbhlr, Salmi_2022_wcauZ}, are also broadly satisfied by our chosen EoS models. As the set spans a wide range of stiffnesses, the tidal deformability constraints from GW170817 \cite{Abbott_2017_TvRVX, Abbott_2018_EX9Nk, Landry_2019_HYTmg, Tews_2018_VRpfs}, which tend to favour softer EoSs, are likewise accommodated.

\subsection{Quark EoSs and Hybrid EoSs}
\label{eos_quark}
In the cores of CSs, matter density significantly exceeds the nuclear saturation density. The QCD phase diagram suggests a transition from hadronic to deconfined quark matter at such high densities \cite{Halasz_1998_ssGjO}. The MIT bag model EoS is widely used to model this transition and study CSs, core-collapse supernovae, etc., involving a quark core \cite{Zdunik_2013_cA65e, _zel_2010_JsOjN, Fischer_2011_SOE3a}. While it offers a simple, parameterizable framework incorporating confinement and asymptotic freedom, it lacks detailed quark interaction treatment at high densities. In contrast, pQCD EoS provides a QCD-consistent, scalable approach, especially where perturbation theory is valid. Lattice QCD studies of hot quark-gluon plasma further highlight the role of interactions in the QCD phase structure. Hence, we adopt the interacting quark matter EoS from Ref. \cite{Fraga_2014_1XoqE}, which accounts for interactions and systematic uncertainties via renormalization scale dependence. This effective EoS, more refined than the MIT model yet comparably tractable, enables stable HS branches, indicating smooth matching with our hadronic EoSs. We apply the Maxwell construction (MC) for the hadron-quark phase transition \cite{Glendenning_2001_zTYrc}, observing a sharp $c_s^2$ change and energy density jump at the common transition pressure. To model possible crossover-like behaviour, we also generate hybrid EoSs using the replacement interpolation method (RIM) from Ref. \cite{Abgaryan_2018_1drtV}.

Further, we utilize the following hybrid EoS tables from the CompOSE database \cite{obspmHome, doiCompOSEReference, Typel_2015_gH0QU}:
(i) RDF(1.1–1.9), generated via a two-phase approach between the DD2F hadronic model and the string-flip quark model \cite{Bastian_2018_K1F9o};
(ii) DD2+QM2 and DD2+QM2+1 from Ref. \cite{Otto_2020_Ldq3I}, constructed by combining the DD2 hadronic model and a quark model based on the functional renormalization group approach with quark-meson truncation, employing the Maxwell phase construction;
(iii) DS(CMF)-2 Hybrid and DS(CMF)-6 Hybrid \cite{obspmHome, Dexheimer_2021_Nc522}, based on the CMF model;
(iv) QHC19(A–D) \cite{Baym_2019_gurlu}, where nucleons are described using the extended APR model \cite{Togashi_2013_eAPR}, deconfined quark matter is modelled via the NJL framework, and the two phases are connected through a smooth interpolation. The resulting mass-radius relations of HSs derived from these hybrid EoSs are consistent with recent astrophysical constraints \cite{Cromartie_2019_NtRiY, Miller_2021_pAcGM, Riley_2019_oQUYh, Riley_2021_kbhlr, Salmi_2022_wcauZ, Tews_2018_VRpfs, Paschalidis_2018_AetFT, Most_2018_JHlGX, Malik_2018_qAfAT, Essick_2020_enQBF, De_2018_LcNIX}.

\section{Equilibrium Configurations of Compact Stars}
\label{equilibrium_model}
In this section, we review the equations to compute observables like mass, radius, tidal deformability, moment of inertia, etc., for static, slowly rotating, and rapidly rotating stars putting $G = c = 1$. Assuming the matter is a perfect fluid, the energy-momentum tensor is

\begin{equation}
T_{\mu \nu} = \left( \varepsilon + P \right) u_{\mu} u_{\nu} + P g_{\mu \nu},
\label{eqn1}
\end{equation}
where $\varepsilon$, $P$, $u_\mu$, and $g_{\mu \nu}$ denote mass-energy density, pressure, fluid four-velocity, and the metric tensor, respectively.

\subsection{Non-rotating Stellar Configurations}

The metric for a static, spherically symmetric star in $(t, r, \theta, \varphi)$ coordinates is

\begin{equation}
ds^2 = - e^{2\nu(r)}dt^2 + e^{2\lambda(r)} dr^2 + r^2(d\theta^2 + \sin^2\theta d\varphi^2),
\label{eqn2}
\end{equation}

with $\nu(r)$ and $\lambda(r)$ are the metric function obtained by sloving Einstein equations. The macroscopic quantities mass and radius are obtained from the solution of Tolman-Oppenheimer-Volkoff (TOV) equations \cite{Tolman_1939_WVTIU, Oppenheimer_1939_oVOhJ}:

\begin{eqnarray}
\frac{dM(r)}{dr} = 4\pi r^2 \varepsilon(r) \\
\frac{dP(r)}{dr} = -\frac{(\varepsilon(r) + P(r))(M(r) + 4\pi r^3 P(r))}{r^2 (1 - \frac{2M(r)}{r})}
\label{eqn3}
\end{eqnarray}
M(r) is the mass enclosed within radius r. Integration of the above equations is performed from $r = 0$ to $r = R$, where $P(R) = 0$. Thus, $R$ defines the surface of the star and $M(R)$ is the total gravitating mass of the static stellar configuration.

\subsection{Slowly Rotating Stars}

Using the Hartle-Thorne formalism \cite{Hartle_1967_7x5QB, Benhar_2005_gKaTj}, the metric for slowly rotating stars is given by

\begin{eqnarray}
& ds^2 = - e^{2\nu\left( r \right)}\left[ 1+2h\left( r, \theta \right) \right]dt^2 + e^{2\lambda(r)} \left[ 1 + \frac{2m\left( r, \theta \right)}{r-2M(r)}\right] dr^2 \nonumber\\
& + r^2 \left[ 1 + 2k\left( r, \theta \right) \right] \left[d\theta^2 + \sin^2\theta \left( d\varphi - \omega\left( r, \theta \right) dt \right)^2 \right],
\label{eqn2a}
\end{eqnarray}

where, $\nu(r)$, $\lambda(r)$ and $M(r)$ represent the background non-rotating configuration. The perturbation functions $h(r, \theta)$, $m(r, \theta)$, $k(r, \theta)$ describe the slowly rotating configuration, and $\omega(r, \theta)$ accounts for the frame dragging. The stars maintain reflection-symmetry with respect to the $\theta = \pi/2$ plane and the perturbation functions are expanded in terms of Legendre polynomials, $P_l\left( \cos\theta \right)$. The $l = 0$ and $l = 2$ components of the Einstein equations, expanded in the order of $\Omega$, are solved alongside the TOV background. The $l = 0$ equations determine the increase in mass due to rotation, while the $l = 2$ equations are solved to obtain the quadrupole moment. The solution of the $t\varphi$ component of the Einstein equation gives the moment of inertia. Here, we consider terms $ > \mathcal{O}(\Omega^2)$ to calculate moment of inertia $I$, angular momentum $J$, and quadrupole moment $Q$ \cite{Benhar_2005_gKaTj, Yagi_2014_lZhJS}.

\subsection{Tidal Deformability}

In the context of stellar non-radial perturbations in very slowly rotating stars, even-parity Regge-Wheeler gauge yields the differential equation for metric perturbation $H_0$ \cite{Thorne_1967_LSquP}:

\begin{eqnarray}
\frac{d^2H_0}{dr^2} + \left\lbrace \frac{2}{ r} + \frac{1}{2} \left( \frac{d\nu}{dr} - \frac{d\lambda}{dr} \right) \right\rbrace \frac{dH_0}{dr} + \nonumber\\
\left\lbrace \frac{2}{r^2} - \left[ 2+l\left(l+1\right) \right] \frac{e^\lambda}{r^2} + \frac{1}{2r} \left( 5\frac{d\lambda}{dr} + 9\frac{d\nu}{dr} \right) \right. \nonumber\\
\left. - \left( \frac{d\nu}{dr} \right)^{2} + \frac{1}{2rc_s^2} \left( \frac{d\nu}{dr} + \frac{d\lambda}{dr} \right) \right\rbrace H_0 = 0
\label{eqn4}
\end{eqnarray}
where $c_s^2 = \frac{dP}{d\varepsilon}$. We apply the boundary conditions outlined in ref. \cite{Hinderer_2008_ZsqiB} to solve this equation for $l=2$. The tidal Love number $k_2$ is given by

\begin{eqnarray}
& k_2 = \frac{8C^5}{5} \left(1 - 2C\right)^2 \left[2 + 2C\left(y - 1\right) - y\right] \nonumber\\
& \times \left\lbrace 2C \left[6 - 3y + 3C(5y - 8) \right] \right. \nonumber\\
& + 4C^3 \left[13 - 11y + C(3y - 2) + 2C^2 (1 + y)\right] \nonumber\\
& \left. + 3(1 - 2C)^2 \left[2 - y + 2C(y - 1)\right] \ln(1 - 2C)\right\rbrace^{-1},
\label{eqn5}
\end{eqnarray}
with $C = M/R$, $y = R \left[ 1/H_0~dH_0/dr \right]_{r=R}$, and dimensionless tidal deformability defined as
\begin{equation}
\overline{\Lambda} = \frac{2}{3} k_2 C^{-5}.
\label{eqn6}
\end{equation}

\subsection{Rapidly Rotating Stars}

The Komatsu-Eriguchi-Hachisu method \cite{Komatsu_1989_hBLFz} provides the framework for axisymmetric, rapidly rotating stars. The metric is
\begin{eqnarray}
ds^2 = & - e^{ \gamma + \rho } dt^2 + e^{2\alpha} \left( dr^2 + r^2 d\theta^2 \right) \nonumber\\
& + e^{ \gamma - \rho } r^2 \sin^2 \theta \left( d\varphi - \omega dt \right)^2,
\label{eqn2b}
\end{eqnarray}

where potentials $\gamma$, $\rho$, $\alpha$, and $\omega$ are functions of $r$ and $\theta$. We employ the numerical technique used in RNS code \cite{Koranda_1997_82RG7, Laarakkers_1999_PPeZV} to obtain $M$, $R$, $I$, $J$, and $Q$.

The dimensionless spin is defined by
\begin{equation}
\chi = \frac{J}{M^2},
\label{eqn7}
\end{equation}

and the scaled moment of inertia and quadrupole moment are
\begin{eqnarray}
\overline{I} &=& \frac{I}{M^3} \nonumber\\
\overline{Q} &=& \frac{Q}{M^3\chi^2}.
\label{eqn8}
\end{eqnarray}

\section{Results and Discussion}
\label{result_discussion}
In this section, we present a detailed analysis of the universal relations among higher-order multipole moments of CSs, such as the moment of inertia and quadrupole moment, for both slowly and rapidly rotating configurations. Our analysis employs a compilation of 22 hadronic EoSs, characterized by a broad range of stiffness properties and internal compositions. The mass--radius constraints from NICER observations \cite{Cromartie_2019_NtRiY, Miller_2021_pAcGM, Riley_2019_oQUYh, Riley_2021_kbhlr, Salmi_2022_wcauZ} as well as the tidal deformability constraints from GW170817 \cite{Abbott_2017_TvRVX, Abbott_2018_EX9Nk, Landry_2019_HYTmg, Tews_2018_VRpfs} are broadly satisfied with this EoS set. We then construct quasi-universal relations for NSs considering the presence of heavy baryons, and delta resonances in the core of the NSs. To isolate the impact of exotic degrees of freedom, such as the presence of heavy baryons in the core, we compare our results with previous investigations \cite{Li_2023_wnUwG, Raduta_2020_YKkHx, Lenka_2019_VAoni, Marques_2017_dQf8a}, highlighting their role in shaping the quasi-universal behaviour. We construct a set of 100 hybrid EoSs using both the Maxwell and Gibbs-like constructions \cite{Bastian_2021_RDF, Baym_2019_gurlu}, and use them to model HS configurations. These HS models are tested against current observational constraints on the mass-radius relation \cite{Cromartie_2019_NtRiY, Miller_2021_pAcGM, Riley_2019_oQUYh, Riley_2021_kbhlr, Salmi_2022_wcauZ}. Additionally, we assess their consistency with tidal deformability and radius measurements derived from multi-messenger observations, particularly GW170817 \cite{Tews_2018_VRpfs, Paschalidis_2018_AetFT, Most_2018_JHlGX, Malik_2018_qAfAT, Essick_2020_enQBF, De_2018_LcNIX}. Finally, we obtain the universal relations for CSs using the combined set of hadronic and hybrid EoSs. To contextualize our findings, we compare our results with earlier studies involving extensive EoS datasets \cite{Suleiman_2021_lhDZ3, Suleiman_2024_7cKmd, Legred_2024_WRfSP, Khosravi_Largani_2022_cvIJr}. Further, we extend our EoS-insensitive relations for CSs using this combined set for arbitrary rotational sequence.

Fig. \ref{fig1H_eos} presents the full set of hadronic EoSs employed in this study, along with their corresponding $c_s^2$ profiles. The complete ensemble, including both hadronic and hybrid EoSs, is displayed in Fig. \ref{fig1A_eos}. For purely nucleonic matter, $c_s^2$ increases monotonically with energy density $\varepsilon$. The inclusion of additional baryonic degrees of freedom reduces $c_s^2$, as illustrated in Fig. \ref{fig1H_eos}(b). The transition of hadronic matter to deconfined quark matter further modifies this behaviour. Under the MC, $c_s^2$ exhibits a discontinuous jump, whereas, in the presence of a mixed phase, the transition appears smooth (Fig. \ref{fig1A_eos}(b)). The onset of quark matter is also accompanied by a notable drop in pressure relative to the hadronic phase. The phase transition, in our HS models, indicates the presence of sizable quark cores in the most massive CS configurations \cite{Annala_2020_6ap28}.

Figure \ref{fig2_M-R} summarizes the mass-radius ($M$-$R$) relations for NSs containing exotic baryons and HSs featuring deconfined quark cores. Panel (a) displays the $M$--$R$ curves for static configurations, overlaid with recent astrophysical constraints. The employed EoS set predicts a maximum mass in the range 1.90 $M_{\odot}$ $\lesssim$ $M_{\text{max}}$ $\lesssim$ 2.75 $M_{\odot}$, and radii for 1.4 $M_{\odot}$ configurations spanning 11.5 $\lesssim$ $R_{1.4}$ $\lesssim$ 14.9 km. These predictions are broadly consistent with NICER measurements for PSRs J0030+0451 and J0740+6620 \cite{Miller_2019_dFNjM, Miller_2021_pAcGM, Riley_2019_oQUYh, Riley_2021_kbhlr}. A minor discrepancy remains as the lower bound of $M_{\text{max}}$ falls short of the value inferred in \cite{Fonseca_2021_SOyMc}. Nonetheless, the presence of deconfined quark matter in HS cores remains compatible with current observational constraints on high-mass pulsars. Among the very stiff hadronic EoSs, DDMEX and NL3 yield maximum masses consistent with the estimated range of the secondary component in GW190814 \cite{Abbott_2020_jIBsD}. However, HS configurations constructed from these EoSs--whether using a sharp first-order transition or a crossover-like transition--fail to reproduce this high mass range. In both phase transition scenarios, the onset of quark matter leads to increased compactness in HS configurations with static masses $\gtrsim$1.4$M_\odot$. On the other hand, low-mass NSs modelled with these stiff EoSs exhibit radii that exceed the constraints inferred for the central compact object within HESS J1731-347, thereby disfavoring such models for such low-mass configurations. 

The $M$-$R$ constraints for the central compact object in HESS J1731--347, derived using parallax priors and X-ray data, are shown in figure \ref{fig2_M-R} for both the 68.3\% and 95.4\% credibility intervals \cite{Doroshenko_2022_xyimd}. Our $M$-$R$ results, computed across the entire EoS ensemble, reproduce the mass range associated with this anomalously low-mass compact object; however, the corresponding radii lie within the 11.4-15.1 km interval, exceeding the estimated value inferred for HESS J1731-347. Among the hadronic EoSs, the nucleonic EoS D1M$^{*}$--based on non-relativistic density functional theory--yields $M$--$R$ predictions that intersect the 68.3\% credibility contour of HESS J1731-347. Within the RMF framework, the DD2Y$\Delta$(b) EoS, which includes the effects of heavy baryons and $\Delta$ resonances, exhibits sufficient softness to produce $M$--$R$ result compatible with the observed constraints. In our HS models, we find that MC between the DD2F hadronic EoS and the string-flip quark EoS results in hybrid EoSs--RDF(1.8) and RDF(1.9) \cite{Bastian_2018_K1F9o}--that generate HS configurations consistent with the HESS J1731--347 $M$--$R$ constraints. Furthermore, the QHC family of hybrid EoSs, incorporating a crossover-like phase transition, also leads to HS models compatible with the constraints. For other HS models in our dataset, the onset of deconfined quark matter occurs beyond the region of the $M$--$R$ plane relevant to HESS J1731--347, rendering them incompatible with this object. Overall, the constraints associated with HESS J1731--347 central compact object are explainable within the EoS set featuring general core compositions. The subset of EoSs that satisfy the $M$--$R$ constraints for this compact object yields a maximum mass in the range 1.93~$M_{\odot}$ $\lesssim$ $M_{\text{max}}$ $\lesssim$2.28$M_{\odot}$, which remains consistent with the other observational constraints from NICER measurements and GW170817.

The Keplerian sequence--depicting the mass--equatorial radius relation for maximally rotating configurations--is shown in Fig.\ref{fig2_M-R}(b). Several EoSs in our dataset yield rotating configurations that surpass the upper mass limit inferred for PSR J0740+6620. The resulting mass range, 2.25$M_{\odot} \leq M_{\text{max}} \leq$ 3.30$M_{\odot}$ for rotational periods $\gtrsim$ 1ms, renders these models particularly relevant for describing rapidly rotating pulsars such as the 2.8ms PSR J0740+6620. Furthermore, with the exception of a few softer EoSs, most models in the EoS ensemble accommodate the mass estimate of the secondary component in GW190814 \cite{Abbott_2020_jIBsD} within their maximally rotating sequences. For such configurations, the predicted equatorial radii at 2 $M_\odot$ span the range 15.9 $\lesssim~R_{2.0}~\lesssim$ 21.1 km, with corresponding spin frequencies between $\sim$800–1200Hz. Across all spin sequences explored in this study, 2 $M_\odot$ CSs exhibit minimum spin frequencies between 320 and 800Hz. These results indicate that the employed EoS dataset is consistent with the observed characteristics of PSR J1748--2446ad \cite{Hessels_2006_3QLXM}. Notably, the emergence of a deconfined quark core leads to a reduction in the maximum mass supported in the Keplerian sequence. This suppression is a direct consequence of the softer high-density behaviour introduced by the quark phase and is accompanied by an increase in both the compactness and the Keplerian (mass-shedding) frequency of the resulting HS configurations, relative to the same mass NS configurations.

\subsection{Relations for slowly rotating compact stars}
\label{RD_slowNS}
The dimensionless tidal Love number, $k_2$, is defined in Eq.\ref{eqn5}, while the dimensionless tidal deformability, $\overline{\Lambda}$, is given by Eq.\ref{eqn6}. The numerical results for $k_2$ are displayed in Fig.~\ref{fig3_k2-C}, which show that $k_2$ decreases monotonically with increasing compactness $C$ for $C \gtrsim 0.1$. However, this trend changes for $C \lesssim 0.1$, a regime particularly relevant for recently inferred low-mass NSs. The Love number $k_2$ depends on both the compactness and the response to tidal perturbations. In this low-compactness regime, the presence of a solid crust reduces the star's deformability, and the observed behaviour is primarily governed by the relatively larger contribution of the crust to the overall structure. On the other hand, in the high-compactness regime, the presence of deconfined quark matter causes the compactness to saturate earlier, typically around $C \approx 0.25$ in static configurations. As a result, compact stars (CSs) tend to cluster near the boundary of the $k_2$–$C$ plane (see Fig.3 of Ref.\cite{Yagi_2014_zajqL}), which increases the likelihood of breaking the emergent symmetry responsible for EoS-insensitive behaviour. From Eq.\ref{eqn6}, one observes that $\overline{\Lambda} \propto k_2 C^{-5}$. Since $k_2$ itself decreases with increasing $C$, the overall scaling implies $\overline{\Lambda} \propto C^{-6}$. This expected trend is visually confirmed in Fig.\ref{fig4_C-L}, where $\overline{\Lambda}$ is plotted on a logarithmic scale. The result clearly demonstrates an EoS-independent relationship between $C$ and $\overline{\Lambda}$. Consequently, the $C$–$\overline{\Lambda}$ relation can be accurately fitted using a polynomial in $\ln \overline{\Lambda}$, following a scaling approach similar to that proposed in Ref.~\cite{Yagi_2013_cPzvP},

\begin{equation}
C = \sum\limits_{n=0}^{m} a_n \left( \ln \overline{\Lambda} \right)^n .
\label{eqn9}
\end{equation}

We fit the $C$-$\overline{\Lambda}$ relation at zero temperature using Eq. \ref{eqn9} with $m=2$ and $m=4$, both yielding reasonable fits for a moderate number of EoSs. We determine the relations separately for the cases of (a) hadronic EoSs only (N, Y, $\Delta$ composition) and (b) hadronic EoSs + hybrid EoSs (N, Y, $\Delta$, q composition). Our results are consistent with earlier findings on the properties of compact stars with exotic degrees of freedom \cite{Li_2019_BXUgg, Li_2020_il2wz}. The relation is also provided in Refs. \cite{Yagi_2017_uWwLw, Li_2023_wnUwG, Raduta_2022_mFVlJ} for $m=2$. The role of non-$\beta$-equilibrated nuclear matter in the relation, considering hadronic EoSs with a few exotic degrees of freedom, is discussed in Ref. \cite{Raduta_2022_mFVlJ}, while Ref. \cite{Li_2023_wnUwG} provides the relation for compact objects with heavy baryons at zero temperature and $\beta$-equilibrium. Comparison of the $m=2$ and $m=4$ results with those from Ref. \cite{Li_2023_wnUwG} reveals an excellent match. We then extend the relation for composition (b) with both $m=2$ and $m=4$. We obtain that $m=4$ yields a slightly better fit in comparison to $m=2$ fitting relation. In Figs. \ref{fig4_C-L}(a) and \ref{fig4_C-L}(b), the fitting functions using $m=4$ in Eq. \ref{eqn9} along with the computed results are shown in the top panels for compositions (a) and (b), respectively. The bottom panels of these figures present the corresponding fractional error between the computed results and the $C$-$\overline{\Lambda}$ fit functions. Our $C$-$\overline{\Lambda}$ fit functions are approximately EoS-independent, with a variation of $\sim 6\%$ for (a) and $\sim 8\%$ for (b). Thus it is noted that the appearance of a quark core tends to further break quasi-universality. The coefficients of the relations for both cases (a) and (b) are given in Table \ref{table1} for both $m=2$ and $m=4$, along with the maximum relative error of the relation, $\Delta_{\text{max}}$. 

This relation has been explored in Refs. \cite{Maselli_2013_KEJBr, Yagi_2017_uWwLw, Raduta_2020_YKkHx, Yeung_2021_dc0JE, Burikham_2022_Kq9sd, Carlson_2023_0TLyw, Li_2023_wnUwG} and several other papers for $m \geq 2$, considering the nucleonic, hadronic, hot nuclear matter, and hybrid EoS models. Results from our calculations are in overall agreement with the $C$-$\overline{\Lambda}$ relations obtained in Refs. \cite{Godzieba_2021_UBDLe, Suleiman_2024_7cKmd}, where the authors used a very large number of EoSs. The relative error for the relation in Ref. \cite{Suleiman_2024_7cKmd}, using an EoS set constructed with the Gaussian process (GP) and constrained from current astrophysical observations, is $\lesssim 12\%$. However, Ref. \cite{Legred_2024_WRfSP} obtained a maximum relative error $\sim 2.0\%$ for the $C$-$\overline{\Lambda}$ relation while introducing phase transition phenomenology.

Next, within the slow rotation approximation we investigate the universal relations associated with the I-Love-Q trio. We explore relations between $\overline{I}$, $\overline{\Lambda}$, and $\overline{Q}$ employing the method introduced in Refs. \cite{Yagi_2013_cPzvP, Yagi_2013_yxUgU},

\begin{equation}
\ln y = \sum\limits_{n=0}^{m} a_n \left( \ln x \right)^n
\label{eqn10}
\end{equation}

where the equation is used for all three pairs of variables ($\overline{\Lambda}$, $\overline{I}$), ($\overline{\Lambda}$, $\overline{Q}$), and ($\overline{Q}$, $\overline{I}$). The quadrupole moment and moment of inertia for slowly rotating stars are calculated using the perturbative approach outlined in Ref. \cite{Benhar_2005_gKaTj}. We take the spin parameter $\chi$ to be a constant ($10^{-3}$) for all CS configurations. Using Eq. \ref{eqn10}, in a log-log scale we take $m=2$, $m=4$ and obtain fit relations for both compositions (a) and (b). We compare our fit relations with existing literature \cite{Baub_ck_2013_gJPHq, Suleiman_2024_7cKmd, Legred_2024_WRfSP, Urbanec_2013_wNmiD, Yeung_2021_dc0JE}. We show the computed results in Figs. \ref{fig5_Ib-L} \& \ref{fig7_Qb-L} along with the fit relations with $m=4$ for ($\overline{\Lambda}$, $\overline{I}$) and ($\overline{\Lambda}$, $\overline{Q}$) variable pairs, respectively. Computed results and fit functions are presented in the upper panels while the fractional differences between the fit relations and numerical results are shown in the bottom panels of the figures. We summarize the fitting parameters along with the maximum relative deviation, $\Delta_{\text{max}}$, in Table \ref{table2}.

The I-Love-Q relations, derived in this work, with composition (a) and $m=4$, are in overall agreement with those provided in Table IV of Ref. \cite{Wang_2022_1ft2d}, which employs the relativistic Brueckner-Hartree-Fock theory in full Dirac space to model NSs. The I-Love-Q relations obtained for NS EoSs with heavy baryon degrees of freedom for spin parameter $\chi \ll 1$ in Ref. \cite{Li_2023_wnUwG} are in good agreement with our obtained results for composition (a). Further, the relations extended for composition (b) are also in reasonable agreement with earlier relations for CSs treated within the slow rotation approximation \cite{Yeung_2021_dc0JE, Burikham_2022_Kq9sd}. Ref. \cite{Legred_2024_WRfSP} studied the same relation with a flexible set of phenomenological non-parametric EoSs by introducing phase transition phenomenology, albeit with a slightly different form for the I-Love-Q relations. Their relations yield approximate universality with $\Delta_{\text{max}} \sim \mathcal{O}(20\%)$, consistent with our calculations summarized in Table \ref{table2}. We notice that $m=4$ yields better-fit relations than those obtained using $m=2$. It must be noted that the maximum deviation for our fit relations stays (a) $\lesssim \mathcal{O}(10\%)$ for purely hadron EoSs with heavy baryon degrees of freedom, (b) $\lesssim \mathcal{O}(20\%)$ for the entire set of EoSs. This suggests that the appearance of deconfined quark degrees of freedom in the core adds further variability to the obtained relations.

Given the universality observed in the $C$--$\overline{\Lambda}$ relation—both in the present work and in several earlier studies \cite{Maselli_2013_KEJBr, Godzieba_2021_UBDLe, Suleiman_2024_7cKmd, Legred_2024_WRfSP}—as well as in the $\overline{I}$--$\overline{\Lambda}$--$\overline{Q}$ relation, it is reasonable to hypothesize a corresponding universality among the $\overline{I}$--$C$--$\overline{Q}$ trio. In this context, we explore the associated universal relations using the functional form proposed in Ref.~\cite{Yagi_2013_yxUgU}:

\begin{equation}
y = \sum\limits_{n=1}^{4} a_n \left( \text{C}^{-1} \right)^n .
\label{eqn11}
\end{equation}

We investigate the $\overline{I}$--$C$ and $\overline{Q}$--$C$ relations for slowly rotating CSs. For a spin parameter of $\chi = 10^{-3}$, we observe EoS-insensitive trends between $\overline{I}$ and $C$ for both compositions (a) and (b). The calculated data, along with the corresponding fit relations using $m = 4$, are shown in the upper panels of Figs.\ref{fig8A_I-C_sr}(a) and \ref{fig8A_I-C_sr}(b) for compositions (a) and (b), respectively. Similarly, Figs. \ref{fig9A_Q-C_sr}(a) and \ref{fig9A_Q-C_sr}(b) present the $\overline{Q}$--$C$ results and the associated quasi-universal fits for $m = 4$. The computed relation between $\overline{I}$ and $\overline{Q}$ at $\chi = 10^{-3}$ is shown in Fig.~\ref{fig6_I-Q_sr}. We summarize the best-fit parameters for the $\overline{I}$--$C$ and $\overline{Q}$--$C$ relations (Eq.\ref{eqn11}), and for the $\overline{I}$--$\overline{Q}$ relation (Eq.\ref{eqn10}) with $m = 4$, in Table~\ref{table3} for both compositions. The lower panels of the corresponding figures display the relative errors between the fitted and numerical results. Maximum deviations for each relation are also reported in Table~\ref{table3}. In our calculations, we find that the $\overline{I}$--$C$--$\overline{Q}$ quasi-universal relations exhibit relatively smaller fractional deviations for core composition (a) compared to composition (b). This indicates that the variability in the universal behaviour increases when the relations are extended to composition (b). For comparison, the $\overline{I}$--$C$ relation reported in Ref.\cite{Suleiman_2024_7cKmd} shows a maximum deviation of approximately 22\% when using their GP+astro EoS set, and about 9\% for other EoS sets. Additionally, comparisons with the relations proposed in Refs.\cite{Baub_ck_2013_gJPHq, Li_2023_wnUwG} for slowly rotating stars reveal good agreement with our results for composition (a). The emergence of such universal relations is attributed to the approximate self-similarity of isodensity contours, which tend to adopt configurations that minimize the system’s energy. This underlying symmetry is further supported by numerical calculations performed in full General Relativity \cite{Yagi_2014_zajqL, Chatziioannou_2014_sPusC}.

\subsection{Relations for rapidly rotating compact stars}
\label{RD_rapidNS}
In this section, we derive EoS-insensitive relations for rapidly rotating CSs involving the $\overline{I}$–$C$–$\overline{Q}$ trio. It has been established in Ref.~\cite{Doneva_2013_8E7jv} that rapid rotation tends to break the universality of the $\overline{I}$–$\overline{Q}$ relation. However, quasi-universal relations can still be constructed along sequences characterized by fixed normalized spin parameters, such as the dimensionless spin $\chi$ \cite{Doneva_2014_K2OBz}. In our calculations, we find that the maximum value of the dimensionless spin parameter, $\chi_{\text{max}}$, corresponding to the Keplerian sequence, remains $\lesssim 0.71$ across the entire EoS set. The inclusion of heavy baryons leads to a reduction in this upper limit in comparison to purely nucleonic EoS \cite{Roy_2025_ur}. We find that the presence of deconfined quark matter further reduces $\chi_{\text{max}}$, reflecting the increased compactness and the softening of the EoSs at high densities.

We then construct the $\overline{I}$–$C$ and $\overline{Q}$–$C$ fits using Eq.\ref{eqn11}, and the $\overline{I}$–$\overline{Q}$ fit using Eq.\ref{eqn10}, with $m = 4$. The resulting $\overline{I}$–$C$ relation for NSs containing heavy baryons, and the corresponding fit across the full EoS dataset, are shown in Figs.~\ref{fig8B_I-C}(a) and \ref{fig8B_I-C}(b), respectively. For each case, results are presented along three sequences corresponding to different values of the spin parameter: $\chi = 0.3$, $\chi = 0.6$, and the Keplerian limit. The upper panels display the numerical data along with the fitted curves, while the lower panels show the relative deviations from the fits. The $\overline{Q}$–$C$ relation, along with the computed data points, is shown in the upper panels of Fig.\ref{fig9B_Q-C} for compositions (a) and (b); the corresponding relative errors appear in the lower panels. Figs.\ref{fig10B_I-Q}(a) and \ref{fig10B_I-Q}(b) display the $\overline{I}$–$\overline{Q}$ relations for NSs with heavy baryonic degrees of freedom in the core and for CSs with general core compositions, respectively. The fit parameters for all rotational sequences, along with the maximum relative deviations $\Delta_{\text{max}}$ between the numerical data and fitted curves, are summarized in Table~\ref{table3}. Thus, quasi-universal relations are applicable even for maximally rotating stars, though the presence of deconfined quarks in the core introduces additional variation.

The study of EoS-insensitive relations remains a topic of significant interest \cite{Rezzolla_2018_Gl7Zu, Kumar_2019_PdDCq}. Accurate measurements of mass in combination with either $\overline{I}$ or $\overline{Q}$ can be used to estimate pulsar radii. However, degeneracies between $\overline{I}$ and $\overline{Q}$—particularly at low spin or high compactness—pose challenges for extracting precise radius estimates from these observables. Although several studies have explored such relations in the slow-rotation regime, investigations under rapid rotation are comparatively limited. For instance, Ref.\cite{Khosravi_Largani_2022_cvIJr} derives $\overline{I}$–$C$ relations for the Keplerian sequence using the framework developed in Ref.\cite{Khadkikar_2021_sIAAy}, incorporating a range of hadronic and quark matter EoSs. These include models based on the MIT bag framework \cite{Fischer_2011_DbzLV}, vector-interaction bag model \cite{Kl_hn_2015_lo3ys}, string-flip model \cite{Bastian_2018_K1F9o}, and NJL-based approaches \cite{Klevansky_1992_2oEgq, Beni__2015_PdKX3, Baym_2018_fVTXR}. Ref.\cite{Breu_2016_WqIRD} examines $\overline{I}$–$C$ relations using 28 nuclear EoSs and the method of Ref.\cite{Lattimer_2005_z0KLn}, reporting maximum relative errors of $\sim$20\%. 

The relation $\overline{I} = \overline{I}(\overline{Q}, \chi)$ for rapidly rotating NSs has been established in Refs.\cite{Pappas_2014_TIiDy, Chakrabarti_2014_kBsdN}, with relative errors $\lesssim$1\% for arbitrary rotation upto mass-shedding limit. More recently, Ref.\cite{Papigkiotis_2023_a8kst} applies machine learning techniques to derive the relation within $\sim$2\% accuracy across EoSs incorporating nucleonic, hyperonic, and deconfined quark matter degrees of freedom. In the present work, we present $\overline{I}$–$C$–$\overline{Q}$ relations for CSs with arbitrary core composition across multiple fixed values of the spin parameter $\chi$. For composition (a), our calculations show that the $\overline{I}$–$C$ quasi-universal relations exhibit relative deviations $\lesssim 10\%$. When extended to composition (b), this quasi-universality is slightly broken, with the maximum relative error, $\Delta_{\text{max}}$, increasing to $\lesssim 15\%$. We also observe a mild increase in $\Delta_{\text{max}}$ with increasing $\chi$. Interestingly, for the Keplerian sequence—which does not correspond to a single fixed value of $\chi$—the maximum relative error is lower than that for the $\chi = 0.6$ sequence. The $\overline{I}$–$C$ relation governed by Eq.~\ref{eqn8} is expected to follow a scaling behaviour of $\propto 1/C^2$. However, for rapidly rotating CS sequences, deviations arise due to two key effects: significant deformation of isodensity surfaces and the additional mass supported by rotation. As a result, we find that deviations are smaller in the high-compactness regime relevant for massive pulsars, and more pronounced in the low-compactness regime relevant for objects such as HESS J1731–347. As $\chi$ increases, the distinction between deviations in the low- and high-compactness regimes diminishes, leading to more uniform behaviour across the full compactness range. The inclusion of quark degrees of freedom in the stellar core introduces additional variability in the quasi-universal relations across all the rotational sequences.

Further, the $\overline{Q}$–$C$ relation also exhibits a decreasing trend with increasing compactness. The quadrupole moment arises from the deformation of isodensity surfaces in stellar configurations. CSs are more susceptible to deformation in the low-compactness regime, leading to larger deviations in this region compared to the high-compactness regime. Similar to the $\overline{I}$–$C$ relation, departure from universal relation becomes uniform across the full compactness range with increasing $\chi$. This indicates the fact that high-density isodensity contours also deform significantly at larger spin values. Further, the presence of deconfined quark degrees of freedom introduces greater variability in the quasi-universal relations. For composition (b), we find a maximum relative deviation of $\Delta_{\text{max}}$ $\approx 15\%$, while for composition (a), $\Delta_{\text{max}}$ $\approx 12\%$ at $\chi = 0.6$. Interestingly, in the Keplerian sequence, the quasi-universal relations for both compositions (a) and (b) exhibit slightly lower $\Delta_{\text{max}}$ values compared to the $\chi = 0.6$ case. Finally, we also obtain EoS-insensitive relations between $\overline{I}$ and $\overline{Q}$, where $\overline{I}$ increases monotonically with $\overline{Q}$. In this relation, the high-$\overline{Q}$ regime (corresponding to low compactness) contributes more to the overall variability. As with the other relations, faster rotation tends to increase deviations in the low-$\overline{Q}$ (high compactness) regime as well. The composition (b) consistently shows greater departures from universality than composition (a) across all the rotational sequences.

\twocolumngrid
\section{Conclusion}
\label{conclusion}
We explored the universal relations for CSs, which allow almost all the exotic degrees of freedom of compressed matter in their cores. We used the hadronic EoSs, consistent with NICER observations of NSs and constraints from nuclear physics experiments. These models predict maximum masses for static configurations in the range 1.95 $M_{\odot}$ $\lesssim$ $M_{\text{max}}$ $\lesssim$ 2.75 $M_{\odot}$, while the radii of 1.4 $M_{\odot}$ static configurations lie within 11.7 km $\lesssim$ $R_{1.4}$ $\lesssim$ 14.9 km. We further considered the possibility of deconfined and interacting quark matter in the cores of CSs. Motivated by lattice QCD studies of the thermodynamics of hot quark-gluon plasma, which highlight the significance of quark interactions, we adopted a framework for interacting quark matter at zero temperature and sufficiently high densities. The hadronic and quark EoSs were joined using different hadron–quark deconfinement phase transition constructions. For our set of hybrid EoSs, the maximum mass of the last stable static configuration satisfies 1.90 $M_{\odot}$ $\lesssim$ $M_{\text{max}}$ $\lesssim$ 2.35 $M_{\odot}$.

We calculate the measurable macroscopic properties of CSs—including mass, radius, tidal deformability, moment of inertia, and quadrupole moment—for isolated configurations using the entire EoS dataset. These include models where the full baryonic octet is present, and the emergence of deconfined quark degrees of freedom at high densities becomes inevitable. The impact of these exotic degrees of freedom on the $C$–$\overline{\Lambda}$ and $\overline{I}$–$\overline{\Lambda}$–$\overline{Q}$ relations for slowly rotating NSs is summarized in Section~\ref{RD_slowNS}. We validate our results by comparison with existing literature on slowly rotating CSs. Subsequently, we utilize the nearly EoS–independent nature of the $C$–$\overline{\Lambda}$ relation to construct the $\overline{I}$–$C$–$\overline{Q}$ relations for rapidly rotating CSs with general core compositions. These relations are investigated across sequences with fixed spin parameters as well as along the maximally rotating Keplerian sequence. Our findings indicate that the fitted relations for each of the rotational sequences are approximately universal, though the appearance of deconfined quark matter in the core increases the associated fractional deviations.

In conclusion, this study investigates the universal relations for CSs under arbitrary rotation, incorporating a broad spectrum of exotic degrees of freedom in nuclear matter at extreme densities. The resulting relations extend and refine those previously reported in the literature, enabling EoS-independent estimates of key macroscopic properties of observed NSs. The synergy between future astrophysical observations and such quasi-universal relations offers a promising avenue for advancing our understanding of ultra-dense matter in compact stellar environments.

\onecolumngrid
\begin{table}[t!]
\centering
\captionsetup{justification=raggedright}
\caption{Fitting parameters for the $C$–$\overline{\Lambda}$ relation obtained using Eq.~(\ref{eqn9}), along with the corresponding maximum relative deviation, $\Delta_{\text{max}}$.}
\label{table1}
\begin{tabularx}{\textwidth}{|p{0.7cm} p{0.7cm} X X X X X X p{1.0cm}|}
\hline
$y$ & $x$ & Comp & $~~a_0$ & $~~a_1$ & $~~a_2$ & $~~~a_3$ & $~~~a_4$ & $\Delta_{\text{max}}$ (\%) \\ 
\hline\hline
\rule{0pt}{3.0ex}
$C$ & $\overline{\Lambda}$ & N, Y, $\Delta$ & $3.6265\times10^{-1}$ & $-3.5618\times10^{-2}$ & $8.5727\times10^{-4}$ & & & 6.3 \\
 &  & N, Y, $\Delta$, q & $3.5760\times10^{-1}$ & $-3.4377\times10^{-2}$ & $7.8159\times10^{-4}$ & & & 8.0 \\
\rule{0pt}{3.0ex}
 &  & N, Y, $\Delta$ & $3.8120\times10^{-1}$ & $-5.1044\times10^{-2}$ & $5.1128\times10^{-3}$ & $-4.7572\times10^{-4}$ & $1.8516\times10^{-5}$ & 5.8 \\
 &  & N, Y, $\Delta$, q & $3.8933\times10^{-1}$ & $-5.9233\times10^{-2}$ & $7.4679\times10^{-3}$ & $-7.3772\times10^{-4}$ & $2.8502\times10^{-5}$ & 7.8 \\
\hline
\end{tabularx}
\end{table}
\begin{table}
\centering
\captionsetup{justification=raggedright}
\caption{Parameters obtained from fitting the $\overline{I}$–$\overline{\Lambda}$, $\overline{Q}$–$\overline{\Lambda}$, and $\overline{I}$–$\overline{Q}$ universal relations using Eq.~(\ref{eqn10}), along with their respective maximum fractional errors, $\Delta_{\text{max}}$.}
\label{table2}
\begin{tabularx}{\textwidth}{|p{0.7cm} p{0.7cm} X X X X X X p{1.0cm}|}
\hline
$y$ & $x$ & Comp & $~~a_0$ & $~~a_1$ & $~~a_2$ & $~~~a_3$ & $~~~a_4$ & $\Delta_{\text{max}}$ (\%) \\
\hline\hline
\rule{0pt}{3.0ex}
$\overline{I}$ & $\overline{\Lambda}$ & N, Y, $\Delta$ & $1.4242\times10^{0}$ & $1.0252\times10^{-1}$ & $1.0607\times10^{-2}$ & & & $7.4$ \\
 &  & N, Y, $\Delta$, q & $1.4622\times10^{0}$ & $9.2261\times10^{-2}$ & $1.1297\times10^{-2}$ & & & $9.9$ \\
\rule{0pt}{3.0ex}
 &  & N, Y, $\Delta$ & $1.4863\times10^{0}$ & $5.3901\times10^{-2}$ & $2.3479\times10^{-2}$ & $-1.3892\times10^{-3}$ & $5.2491\times10^{-5}$ & $7.3$ \\
 &  & N, Y, $\Delta$, q & $1.2917\times10^{0}$ & $2.2566\times10^{-1}$ & $-2.4984\times10^{-2}$ & $4.0916\times10^{-3}$ & $-1.6312\times10^{-4}$ & $9.7$ \\

\hline
\rule{0pt}{3.0ex}
$\overline{Q}$ & $\overline{\Lambda}$ & N, Y, $\Delta$ & $-7.1640\times10^{-2}$ & $2.7182\times10^{-1}$ & $8.3009\times10^{-4}$ & & & $7.7$ \\
 &  & N, Y, $\Delta$, q & $-3.8118\times10^{-2}$ & $2.6430\times10^{-1}$ & $1.2604\times10^{-3}$ & & & $13.5$ \\
 \rule{0pt}{3.0ex}
 &  & N, Y, $\Delta$ & $1.3612\times10^{-1}$ & $1.1985\times10^{-1}$ & $3.7643\times10^{-2}$ & $-3.5522\times10^{-3}$ & $1.1680\times10^{-4}$ & $7.5$ \\
 &  & N, Y, $\Delta$, q & $-1.3823\times10^{-1}$ & $3.6292\times10^{-1}$ & $-3.0986\times10^{-2}$ & $4.2200\times10^{-3}$ & $-1.8986\times10^{-4}$ & $13.2$ \\

\hline
\rule{0pt}{3.0ex}
$\overline{I}$ & $\overline{Q}$ & N, Y, $\Delta$ & $1.4625\times10^{0}$ & $3.8626\times10^{-1}$ & $1.3520\times10^{-1}$ & & & $1.7$ \\
 &  & N, Y, $\Delta$, q & $1.4772\times10^{0}$ & $3.7013\times10^{-1}$ & $1.3919\times10^{-1}$ & & & $2.8$ \\
 \rule{0pt}{3.0ex}
 &  & N, Y, $\Delta$ & $1.4047\times10^{0}$ & $5.1619\times10^{-1}$ & $5.2583\times10^{-2}$ & $1.4195\times10^{-2}$ & $5.6662\times10^{-4}$ & $1.0$ \\
 &  & N, Y, $\Delta$, q & $1.3981\times10^{0}$ & $5.3742\times10^{-1}$ & $3.0988\times10^{-2}$ & $2.2826\times10^{-2}$ & $-6.2051\times10^{-4}$ & $1.5$ \\
\hline
\end{tabularx}
\end{table}
\onecolumngrid
\begin{table}[t!]
\centering
\captionsetup{justification=raggedright}
\caption{Fitting parameters for the $\overline{I}$–$C$ and $\overline{Q}$–$C$ relations using Eq.(\ref{eqn11}), and for the $\overline{I}$–$\overline{Q}$ relation using Eq.(\ref{eqn10}), along with the corresponding maximum relative deviations, $\Delta_{\text{max}}$.}
\label{table3}
\begin{tabularx}{\textwidth}{|p{0.6cm} p{0.4cm} p{1.5cm} p{1.5cm} X X X X X p{0.7cm}|}
\hline
$~y$ & $x$ & Comp & $\chi$ & $~~a_0$ & $~~a_1$ & $~~a_2$ & $~~~a_3$ & $~~~a_4$ & $\Delta_{\text{max}}$ (\%) \\
\hline\hline
\rule{0pt}{3.0ex}
$\overline{I}$ & $C$ & N, Y, $\Delta$ & slow & & $7.3515\times10^{-1}$ & $2.3983\times10^{-1}$ & $-1.9005\times10^{-4}$ & $-1.8861\times10^{-4}$ & $5.0$ \\
 &  & N, Y, $\Delta$, q & & & $7.3668\times10^{-1}$ & $2.3848\times10^{-1}$ & $-1.5768\times10^{-4}$ & $-1.8253\times10^{-4}$ & $6.5$ \\
 \rule{0pt}{3.0ex}
 &  & N, Y, $\Delta$ & $0.3$ & & $7.2176\times10^{-1}$ & $2.4314\times10^{-1}$ & $-1.1255\times10^{-3}$ & $-1.5455\times10^{-4}$ & $5.0$ \\
 &  & N, Y, $\Delta$, q & & & $7.3174\times10^{-1}$ & $2.3810\times10^{-1}$ & $-5.9093\times10^{-4}$ & $-1.6872\times10^{-4}$ & $6.8$ \\
 \rule{0pt}{3.0ex}
 &  & N, Y, $\Delta$ & $0.4$ & & $7.1656\times10^{-1}$ & $2.4098\times10^{-1}$ & $-1.1819\times10^{-3}$ & $-1.5983\times10^{-4}$ & $5.1$ \\
 &  & N, Y, $\Delta$, q & & & $7.3332\times10^{-1}$ & $2.3403\times10^{-1}$ & $-4.9276\times10^{-4}$ & $-1.7765\times10^{-4}$ & $7.0$ \\
 \rule{0pt}{3.0ex}
 &  & N, Y, $\Delta$ & $0.6$ & & $6.7782\times10^{-1}$ & $2.3468\times10^{-1}$ & $-2.3281\times10^{-3}$ & $-1.2477\times10^{-4}$ & $7.2$ \\
 &  & N, Y, $\Delta$, q & & & $7.2454\times10^{-1}$ & $2.1617\times10^{-1}$ & $-3.3935\times10^{-4}$ & $-1.8667\times10^{-4}$ & $11.0$ \\
 \rule{0pt}{3.0ex}
 &  & N, Y, $\Delta$ & Kep. & & $7.0625\times10^{-1}$ & $1.8221\times10^{-1}$ & $-3.2628\times10^{-3}$ & $-1.2642\times10^{-5}$ & $5.4$ \\
 &  & N, Y, $\Delta$, q & & & $7.3238\times10^{-1}$ & $1.7374\times10^{-1}$ & $-2.5395\times10^{-3}$ & $-2.9869\times10^{-5}$ & $7.1$ \\

\hline
\rule{0pt}{3.0ex}
$\overline{Q}$ & $C$ & N, Y, $\Delta$ & slow & & $-5.7026\times10^{-1}$ & $3.8261\times10^{-1}$ & $-2.4933\times10^{-2}$ & $5.8209\times10^{-4}$ & $10.0$ \\
 &  & N, Y, $\Delta$, q & & & $-5.7609\times10^{-1}$ & $3.8422\times10^{-1}$ & $-2.5153\times10^{-2}$ & $5.9269\times10^{-4}$ & $13.0$ \\
 \rule{0pt}{3.0ex}
 &  & N, Y, $\Delta$ & $0.3$ & & $-5.0184\times10^{-1}$ & $3.2861\times10^{-1}$ & $-1.9511\times10^{-2}$ & $4.0467\times10^{-4}$ & $7.3$ \\
 &  & N, Y, $\Delta$, q & & & $-4.6589\times10^{-1}$ & $3.1346\times10^{-1}$ & $-1.7616\times10^{-2}$ & $3.3138\times10^{-4}$ & $13.4$ \\
 \rule{0pt}{3.0ex}
 &  & N, Y, $\Delta$ & $0.4$ & & $-4.3613\times10^{-1}$ & $2.9452\times10^{-1}$ & $-1.7119\times10^{-2}$ & $3.5233\times10^{-4}$ & $8.8$ \\
 &  & N, Y, $\Delta$, q & & & $-4.2801\times10^{-1}$ & $2.9108\times10^{-1}$ & $-1.6752\times10^{-2}$ & $3.4133\times10^{-4}$ & $11.2$ \\
 \rule{0pt}{3.0ex}
 &  & N, Y, $\Delta$ & $0.6$ & & $-2.3249\times10^{-1}$ & $1.9637\times10^{-1}$ & $-9.5884\times10^{-3}$ & $1.5641\times10^{-4}$ & $11.9$ \\
 &  & N, Y, $\Delta$, q & & & $-2.2562\times10^{-1}$ & $1.9301\times10^{-1}$ & $-9.2043\times10^{-3}$ & $1.4442\times10^{-4}$ & $14.9$ \\
 \rule{0pt}{3.0ex}
 &  & N, Y, $\Delta$ & Kep. & & $-8.3873\times10^{-2}$ & $1.2489\times10^{-1}$ & $-5.6548\times10^{-3}$ & $9.5793\times10^{-5}$ & $6.9$ \\
 &  & N, Y, $\Delta$, q & & & $-8.6403\times10^{-2}$ & $1.2563\times10^{-1}$ & $-5.7326\times10^{-3}$ & $9.8538\times10^{-5}$ & $9.2$ \\

\hline
\rule{0pt}{3.0ex}
$\overline{I}$ & $\overline{Q}$ & N, Y, $\Delta$ & slow & $1.4047\times10^{0}$ & $5.1619\times10^{-1}$ & $5.2583\times10^{-2}$ & $1.4195\times10^{-2}$ & $5.6662\times10^{-4}$ & $1.0$ \\
 &  & N, Y, $\Delta$, q & & $1.3981\times10^{0}$ & $5.3742\times10^{-1}$ & $3.0988\times10^{-2}$ & $2.2826\times10^{-2}$ & $-6.2051\times10^{-4}$ & $1.5$ \\
 \rule{0pt}{3.0ex}
 &  & N, Y, $\Delta$ & $0.3$ & $1.5066\times10^{0}$ & $4.4610\times10^{-1}$ & $1.0763\times10^{-1}$ & $1.5191\times10^{-4}$ & $1.3267\times10^{-3}$ & $7.3$ \\
 &  & N, Y, $\Delta$, q & & $1.5042\times10^{0}$ & $4.5707\times10^{-1}$ & $9.5372\times10^{-2}$ & $4.9758\times10^{-3}$ & $7.2517\times10^{-4}$ & $19.5$ \\
 \rule{0pt}{3.0ex}
 &  & N, Y, $\Delta$ & $0.4$ & $1.5049\times10^{0}$ & $5.0385\times10^{-1}$ & $9.6473\times10^{-2}$ & $-2.8133\times10^{-3}$ & $2.2597\times10^{-3}$ & $1.7$ \\
 &  & N, Y, $\Delta$, q & & $1.5036\times10^{0}$ & $5.1197\times10^{-1}$ & $8.5683\times10^{-2}$ & $2.1872\times10^{-3}$ & $1.5028\times10^{-3}$ & $3.8$ \\
 \rule{0pt}{3.0ex}
 &  & N, Y, $\Delta$ & $0.6$ & $1.5041\times10^{0}$ & $6.5127\times10^{-1}$ & $4.1717\times10^{-2}$ & $9.2124\times10^{-3}$ & $7.3800\times10^{-4}$ & $1.1$ \\
 &  & N, Y, $\Delta$, q & & $1.5030\times10^{0}$ & $6.6043\times10^{-1}$ & $2.7224\times10^{-2}$ & $1.6861\times10^{-2}$ & $-5.5462\times10^{-4}$ & $1.7$ \\
 \rule{0pt}{3.0ex}
 &  & N, Y, $\Delta$ & Kep. & $1.5137\times10^{0}$ & $6.6869\times10^{-1}$ & $8.5352\times10^{-2}$ & $-1.1656\times10^{-2}$ & $2.8662\times10^{-3}$ & $3.6$ \\
 &  & N, Y, $\Delta$, q & & $1.5124\times10^{0}$ & $6.9240\times10^{-1}$ & $4.1866\times10^{-2}$ & $1.2455\times10^{-2}$ & $-1.2613\times10^{-3}$ & $3.6$ \\
\hline
\end{tabularx}
\end{table}

\newpage
\twocolumngrid
\bibliography{pbib}
\end{document}